\documentstyle[12pt,amssymb,aps,epsf]{revtex}

\def\JSP#1#2#3{{\sl J. Stat. Phys.} {\bf #1} (#2) #3}
\def\PRL#1#2#3{{\sl Phys. Rev. Lett.} {\bf#1} (#2) #3}

\def\EPL#1#2#3{{\sl Europhys. Lett.} {\bf#1} (#2) #3}
\def\NPB#1#2#3{{\sl Nucl. Phys.} {\bf B#1} (#2) #3}

\def\CMP#1#2#3{{\sl Comm. Math. Phys.} {\bf #1} (#2) #3}

\def\PRA#1#2#3{{\sl Phys. Rev.} {\bf A#1} (#2) #3}

\def\PRE#1#2#3{{\sl Phys. Rev.} {\bf E#1} (#2) #3}

\def\APNY#1#2#3{{\sl Ann. Phys. (N.Y.)} {\bf #1} (#2) #3}

\def\JP#1#2#3{{\sl J. Physique} {\bf I#1} (#2) #3}
\def\JPA#1#2#3{{\sl J. Physics} {\bf A#1} (#2) #3}

\def\JSM#1#2#3{{\sl J. Soviet Math.} {\bf #1} (#2) #3}

\begin{document}
\def\be{\begin{equation}}
\def\ee{\end{equation}}
\def\bea{\begin{eqnarray}}
\def\eea{\end{eqnarray}}
\def\g{\gamma}
\def\b{\beta}
\def\d{\delta}
\def\a{\alpha}
\def\e{\varepsilon}
\def\la{\lambda}
\def\La{\Lambda}
\def\nn{\nonumber\\}
\def\G{\Gamma}
\def\w{\langle W|}
\def\tA{\tilde A}
\def\tB{\tilde B}
\def\v{|V\rangle}
\def\wt{\langle {\tilde W}|}
\def\vt{|{\tilde V}\rangle}
\def\r#1{(\ref{#1})}
\def\s{\sqrt{\frac{p}{q}}}
\def\sm{{\bar Q}}
\def\cq{{\cal Q}}
\def\t#1{\langle\tau_{#1}\rangle}
\def\2t#1#2{\langle\tau_{#1}\tau_{#2}\rangle}
\def\up{\uparrow}
\begin{titlepage}
\thispagestyle{empty}
\begin{center}
\vspace*{1cm}
{\Large\sc Representations of the quadratic Algebra and\\
Partially Asymmetric Diffusion with Open Boundaries}\\[10mm]
{ {\sc Fabian H.L. E\char'31ler}\footnote{%
e-mail: {\tt fab@thphys.ox.ac.uk}}
{\sc and Vladimir Rittenberg} \footnote{%
e-mail: {\tt unp01c@ibm.rhrz.uni-bonn.de}}}\\[9mm]

\begin{center}
{\small\sl
Physikalisches Institut der Universit\"at Bonn\\
Nussallee 12\\
53115 Bonn, GERMANY}
\end{center}
\vspace*{1cm}
ABSTRACT
\end{center}

\baselineskip=14pt

We consider the one-dimensional partially asymmetric exclusion model
with open boundaries. The model describes a system of hard-core
particles that hop stochastically in both directions with different
rates. At both boundaries particles are injected and extracted.
By means of the method of Derrida, Evans, Hakim and Pasquier the
stationary probability measure can be expressed as a matrix-product
state involving two matrices forming a Fock-like representation of a
general quadratic algebra. We obtain the representations of this
algebra, which were unknown in the mathematical literature and use the
two-dimensional one to derive exact expressions for the density
profile and correlation functions. Using the correspondence between
the stochastic model and a quantum spin chain, we obtain exact
correlation functions for a spin-$\frac{1}{2}$ Heisenberg XXZ chain
with non-diagonal boundary terms. Generalizations to other
reaction-diffusion models are discussed.  
\vspace*{.5cm}

\noindent
{\em PACS numbers:} 05.40+j, 05.60.+w, 64.60, 75.10J\hfill
\end{titlepage}
\baselineskip=14pt
\section{Introduction}
One dimensional reaction-diffusion processes recently have attracted 
much attention for a variety of reasons. Pure diffusion models have been
studied in relation with interface growth \cite{growth}, traffic-flow
\cite{traffic}, the dynamics of shocks \cite{shocks,lebo}, and
magnetophoresis of tagged polymers \cite{poly}.
More general reaction-diffusion models are of interest from a mathematical
point of view due to their relation to integrable quantum chain
hamiltonians \cite{vlad1}. It is interesting to note the important
role played by the boundary conditions in these models \cite{krug}, which
completely control the physics in some cases. 
For the case of two-state models for example, in the corresponding
quantum chain hamiltonians (which are XXZ models) the boundary
conditions generally break the particle number $U(1)$ symmetry and are
not easily treatable by the usual methods like the Bethe Ansatz
\cite{BA1,BA}. The problem is that although the chains can be shown to
be integrable \cite{deV}, the Bethe Ansatz so far has not been
constructed due to the lack of a reference state. 

An important step forward in these types of problems was made by
Derrida, Domany and Mukamel \cite{ddm} in the case of
completely asymmetric diffusion with particle injection at one end of
the chain, and particle extraction at the other end of the chain. They
showed that there exists a recursion, which relates the probability
distribution of the steady state for a $L$ sites to the one for $L-1$
sites. An equivalent formulation of this property was given by
Derrida, Evans, Hakim and Pasquier (DEHP)\cite{dehp}, who demonstrated
that the probability distribution can be written in a factorized form
with coefficients that are not c-numbers but (infinite-dimensional)
matrices. 
For the two-state model there are two matrices which form a Fock
representation of the quadratic algebra. Using representations of this
algebra one can compute in principle all correlation functions. 
In particular the density profile was determined in \cite{dehp,sd},
and in a special case (when the injection rate is equal to the
extraction rate) even the two-point correlation function \cite{de} was
obtained. If one considers the more general problem with particle
injection and extraction at both ends, and partially asymmetric
diffusion, the DEHP approach is still applicable, but the
representations of the quadratic algebra were not known. In a
remarkable paper \cite{sandow}, Sandow was able to compute some
important matrix elements in the envelopping algebra, which allowed
him to compute the currents and to obtain the phase diagram, which
coincides with the mean-field predictions.\vskip .5cm 

In the present paper we start by determining all Fock representations
of the most general quadratic algebra, which depends on seven
parameters (section II). They might be of interest in other physical
contexts as well. It turns out that the representations can be
either finite-dimensional or infinite-dimensional. For each
finite-dimensional representation one obtains a constraint equation
for the seven parameters. This constraint depends on the dimension of
the representation.
The matrix elements of the two matrices appearing in the quadratic
algebra are given by recursion relations. We show that for some
special cases these recursions can be easily solved.

Next we review the connection between the steady-state probability
distribution and the ground state of certain (in general
non-hermitian) quantum chains (section III). In section IV we give a
summary of the DEHP Ansatz and establish the connection with the
quadratic algebra discussed in section II. In section V we consider
the most general master equation for one-dimensional systems with
two-body interactions (one has twelve independent rates) and particle
injection and extraction at both boundaries and apply the DEHP Ansatz. 
One gets a quadratic algebra and two additional quadratic relations on
the matrices. This implies that only finite-dimensional
representations have a chance to be useful. We found that there exists
a one-dimensional representation with three conditions for the twelve
rates and the four parameters describing the injection and extraction
of particles at the ends of the chain. The question of the existence
of higher-dimensional representations and their physical relevance is
left open. In Appendix B we study the applicability of the DEHP
formalism to the steady state of a master equation with three-body
interactions. We show that is this case one obtains, as expected,
cubic algebras. Their representations and physical interest remains to
be studied. 

After these mathematical investigations we turn to a detailed study of
the problem of partially asymmetric diffusion with particle
injection/extraction at both boundaries. We start with a review of the
known results in section VI, using the phase diagram obtained by
Sandow \cite{sandow} as a basis. In the following sections we
concentrate on the application of the two-dimensional representation
of the quadratic algebra to concrete calculations. In sections VII and
VIII we show that as a result of the constraint equation for the
existence of the representation one can cover parts of the phases
$A_{II}$, $B_{II}$, the complete phases $A_{I}$ and $B_{I}$ as well
as the coexistence line (in terms of the definitions of Sch\"utz and
Domany\cite{sd}). The calculation of the density profiles and two-point
correlation functions in the low- and high-density phases is
presented in section IX. One remarkable result is that the density
around the center of the chain has a simple expression in terms of the
parameters of the problem and that it coincides with the mean-field
results (which are derived in Appendix F). The density profile and
two-point function on the coexistence line are presented in section X.

Some by-products of our investigations are presented in the
appendices. In Appendix C we give some identities concerning
normal-ordered expressions of q-oscillators. In Appendix D we show how
the DEHP Ansatz can be used to construct irreducible representations
of the quantum group $U_q(SU(2))$. We close with a discussion of our
results and some remarks on the time-dependence of correlation
functions. 

\section{Fock Representations of the quadratic Algebra}
We are interested in Fock representations of the most general
quadratic algebra
\bea
&&x_1 A^2+x_2AB+x_3BA+x_4B^2=x_5A+x_6B+x_7\ ,\\
\label{ab}
&&A\v\ =0\ ,\quad \w B=0\ ,\quad \langle W|V\rangle\neq 0\ .
\label{ab2}
\eea
Here $x_i$ are complex parameters and quantities of physical interest
are given by vacuum average values of monomials written in terms of
$A$ and $B$, {\sl e.g.}
\be
\w A^{r_1}B^{r_2}\ldots B^{r_n}\v\ .
\label{vev}
\ee
Obviously \r{ab} generalizes the algebra of creation and annihilation
operators and its q-deformations. As far as we know the problem
formulated above has not been considered in the mathematical
literature (presumably because up to now there was no motivation to do
so). General quadratic algebras were studied in \cite{words} but no
Fock-representations were considered.
We will show in the present work that solving the above problem
allows for the computation of concentration profiles and various
correlation functions in the physical problem of partially asymmetric
diffusion with open boundaries.

One can ask the question about the conditions on the $x_i$'s in
\r{ab} such that algebra determines the vacuum expectation value \r{vev}.
This implies that the system of equations for words of length two,
three, {\sl etc} have solutions. Direct calculations show that an
infinite set of inequalities have to be satisfied
\be
x_2\neq 0\ ,\quad x_1x_4-x_2^2\neq 0\ ,\quad
x_2(x_2^2-x_1x_4)+(x_3-x_2)x_1x_4\neq 0\ ,\quad {\sl etc}\ .
\label{vier}
\ee
In section IV we will give, in a different parametrization, a simpler
expression for these conditions. Instead of solving linear equations
for words of different lengths it is useful to look for matrix
representations of the algebra. Once those are known the calculation
of the quantities \r{vev} is simple. We are interested for obvious
reasons in the representations of the smallest dimension because this
is sufficient to compute the relevant quantities. As we are going to
show the representations of the quadratic
algebra are infinite-dimensional unless there exist certain
constraints on the parameters $x_i$. The simplest such constraint is
$x_7=0$, for which the representation is one-dimensional: $A=B=0$.
From now on we will take $x_7\neq 0$. First we consider the case
$x_5\neq0\neq x_6$. We then define
\be
\tA=\frac{x_5}{x_7}A\ ,\quad \tB=\frac{x_6}{x_7}B\ ,
\label{bb}
\ee
in terms of which the algebra reads
\be
z_1 \tA^2+z_2\tA\tB+z_3\tB\tA+z_4\tB^2=\tA+\tB+1\ ,
\label{cc}
\ee
where the $z_i$ are given in terms of the $x_j$ and where
\be
\w \tB=\tA\v\ =0,\quad \langle W|V\rangle\neq 0\ .
\ee
It is convenient to define
\be
\xi=-\frac{z_1}{z_2}=-\frac{x_1x_6}{x_2x_5}\ ,\ \la=-\frac{z_3}{z_2}
=-\frac{x_3}{x_2}\ ,\ \eta=-\frac{z_4}{z_2}=-\frac{x_4x_5}{x_2x_6 }\
,\ z_2=\frac{x_2x_7}{x_5x_6}\ . 
\label{lex}
\ee
One can show (see Appendix A) that through a similarity transformation
the matrices $\tA$ and $\tB$ can be brought to a tridiagonal form with
$\v=\pmatrix{1\cr 0\cr \vdots\cr 0}$ and $\w= \pmatrix{1& 0&\vdots &
0}$. Using this fact together with equation \r{cc} we obtain

\be
\tA=\left(\matrix{a_1&f_1&0&0&0&0&0\ldots\cr
\a_1\eta f_1&a_2&f_2&0&0&0&0\ldots\cr
0&\a_2\eta f_2&a_3&f_3&0&0&0\ldots\cr
0&0&\a_3\eta f_3&a_4&f_4&0&0\ldots\cr
0&0&0&\a_4\eta f_4&a_5&f_5&0\ldots\cr
\ldots& & & & \ldots& &\cr}\right)
\label{a}
\ee

\be
\tB=\left(\matrix{b_1&\a_1\xi f_1&0&0&0&0&0\ldots\cr
f_1&b_2&\a_2\xi f_2&0&0&0&0\ldots\cr
0&f_2&b_3&\a_3\xi f_3&0&0&0\ldots\cr
0&0&f_3&b_4&\a_4\xi f_4&0&0\ldots\cr
0&0&0&f_4&b_5&\a_5\xi f_5&0\ldots\cr
\ldots& & & &\ldots& &\cr}\right)\ .
\label{b}
\ee
The quantities $\a_n,a_n,b_n$ and $f_n$ are given recursively. First
the $\a_n$'s are to be determined from
\be
\a_n=\frac{1+\la\a_{n-1}}{1-\eta\xi\a_{n-1}}\ ,\ \a_1=0\ .
\label{rec1}
\ee
Next one determines $a_n$ and $b_n$ from
\bea
\pmatrix{a_{n+1}\cr b_{n+1}}&=& M\left[\left(\matrix{
\xi(1-\a_n)&\la+\eta\xi\a_n\cr \la+\eta\xi\a_n & \eta(1-\a_n)\cr}\right)
\pmatrix{a_n\cr b_n}+\frac{1}{z_2}\pmatrix{1+\xi\a_n\cr 1+\eta\a_n\cr}
\right]\ ,\nn
M&=&\frac{1}{\eta\xi(1+\la\a_n)^2-(1-\eta\xi\a_n)^2}
\left(\matrix{-\eta(1+\la\a_n)&-1+\eta\xi\a_n\cr
-1+\eta\xi\a_n & -\xi(1+\la\a_n)\cr}\right) \ ,\nn
a_1&=&0=b_1\ .
\label{rec2}
\eea
Finally the $f_n$'s are then given as
\bea
f_n^2&=&f_{n-1}^2\frac{\la+2\eta\xi\a_{n-1}-\eta\xi \a_{n-1}^2}
{1-2\eta\xi\a_n-\la\eta\xi\a_n^2}+\frac{\xi a_n^2+\eta b_n^2+(\la
-1)a_nb_n+\frac{a_n+b_n+1}{z_2}}{1-2\eta\xi\a_n-\la\eta\xi\a_n^2}\
,\nn 
f_0&=&0\ .
\label{rec3}
\eea
For later use we give the first few values
\bea
\a_2&=&1\ ,\ \a_3=\frac{1+\la}{1-\eta\xi}\ ,\nn
a_2&=&\frac{1+\eta}{z_2(1-\eta\xi)}\ ,\
b_2=\frac{1+\xi}{z_2(1-\eta\xi)}\ ,\nn
f_1^2&=&\frac{1}{z_2}\ ,\
f_2^2=\frac{\la+1}{z_2^2(1-2\eta\xi-\la\eta\xi)}
\left[z_2 +\frac{(1+\xi)(1+\eta)}{(1-\eta\xi)^2}\right].
\label{ini}
\eea
We note that for $\la=-1$ it follows that $f_2=0$ and we thus obtain
{\sl only two-dimensional representations} like for one fermion
(observe the appearance of an anticommutator in \r{cc}).
The recursion \r{rec1} for $\a_n$ can be solved by redefining 
\be
\a_n=\frac{1}{\eta\xi}\left(1+\sqrt{\la+\eta\xi}
\frac{u_{n+1}}{u_n}\right), 
\ee
where the $u_n$'s satisfy the following recursion relation
\bea
u_{n+1}+u_{n-1}&=&-\frac{1+\la}{\sqrt{\la+\eta\xi}}u_n\ ,\nn
u_1&=&1\ ,\quad u_2= -\frac{1}{\sqrt{\la+\eta\xi}}\ .
\eea
This is recognized as the special case $x=-\frac{1+\la}{2(\la +\eta\xi)}$
of the recursion relation for Chebyshev polynomials
$U_{n+1}(x)+U_{n-1}(x)=2x U_{n}(x)$. Using the representation
$U_n(x)=\frac{\sin((n-1)\arccos(x)+\phi)}{\sin(\phi)}$ and taking into
account the initial conditions we arrive at the result
\bea
\a_n&=&\frac{1}{\eta\xi}\left[1+\sqrt{\la+\eta\xi}\
\frac{\sin\left((n-1)\theta + \phi\right)}
{\sin\left((n-2)\theta+\phi\right)}\right]\ ,\nn
\phi&=&\arctan\left[
\frac{\sqrt{4\eta\xi-(1-\la)^2}}{\la-1}\right]\ , \
\theta=\arccos\left[-\frac{1+\la}{2\sqrt{\la+\eta\xi}}\right]\ .
\label{alphan}
\eea
The recursion \r{rec2} for $a_n$ and $b_n$ can be decoupled into
recursion relations for ${\tilde a}_n=\sqrt{\xi}a_n+\sqrt{\eta}b_n$
and ${\tilde b}_n=\sqrt{\xi}a_n-\sqrt{\eta}b_n$:
\be
{\tilde a}_{n+1}= g^+_{n+1} {\tilde a}_n +h^{+}_{n+1}\ ,\ 
{\tilde b}_{n+1}= g^{-}_{n+1} {\tilde b}_n +h^{-}_{n+1}\ ,\ 
\ee
where $h^\pm_{n+1}=c_1\pm c_2$, $g^\pm_{n+1}=c_3\pm c_4$, and
\bea
c_1&=&-\sqrt{\xi}\frac{\eta(1+\la\a_n)(1+\xi\a_n)+(1-\eta\xi\a_n)
(1+\eta\a_n)}{z_2(\eta\xi(1+\la\a_n)^2-(1-\eta\xi\a_n)^2)}\,\nn
c_2&=&-\sqrt{\eta}\frac{(1+\xi\a_n)(1-\eta\xi\a_n)+\xi(1+\eta\a_n)
(1+\la\a_n)}{z_2(\eta\xi(1+\la\a_n)^2-(1-\eta\xi\a_n)^2)}\,\nn
c_3&=&-\frac{(\la+\eta\xi)(1-\eta\xi\a_n^2)}
{\eta\xi(1+\la\a_n)^2-(1-\eta\xi\a_n)^2}\ ,\nn
c_4&=&-\frac{\sqrt{\eta\xi}(\la+1)(1+(\la-1)\a_n+\eta\xi\a_n^2)}
{\eta\xi(1+\la\a_n)^2-(1-\eta\xi\a_n)^2}\ .
\label{cs}
\eea
It is hard to simplify the recursion relations further.
Using the expression \r{alphan} for the $\a_n$'s and \r{cs} one can
derive formulae for $a_n$, $b_n$ and hence $f_n$. The resulting
expressions are obviously very cumbersome.
From the expressions \r{a} and \r{b} for $\tA$ and $\tB$ it follows
that the condition for having an $n$-dimensional representation is
simply $f_n=0$. As one can see from the form of the recurrence
relations this constraint is a complicated function of $\la,\xi,\eta$
and $z_2$. This is the reason why we will use for applications only
the two-dimensional representation ($f_2=0$), for which the matrix
elements of $\tA$ and $\tB$ are given by \r{ini}.\vskip .5cm

Let us now consider the case $x_5=x_6=0$. The cases where only $x_5$
or $x_6$ vanishes can be studied in a similar way and thus will not be
considered in detail here. The algebra for case $x_5=x_6=0$ is
\be
z_1A^2+ z_2AB+ z_3BA+ z_4B^2=1\ ,
\label{quads1}
\ee
where $z_i=\frac{x_i}{x_7}$. We define, like in \r{lex}
\be
\xi=-\frac{z_1}{z_2}\ ,\ \la=-\frac{z_3}{z_2}\ ,\
\eta=-\frac{z_4}{z_2}\ .
\label{exl2}
\ee
The infinite-dimensional representation of \r{quads1} is of the form
\r{a}-\r{b} with vanishing diagonal terms $a_n=b_n=0\ ,\forall n$,
where $\eta,\xi,\la$ are defined in \r{exl2} and where $\a_n$ and
$f_n$ are given by
\bea
\a_n&=&\frac{1+\la\a_{n-1}}{1-\eta\xi\a_{n-1}}\ ,\ \a_1=0\ ,\nn
f_n^2&=&f_{n-1}^2\frac{\la+2\eta\xi\a_{n-1}-\eta\xi \a_{n-1}^2}
{1-2\eta\xi\a_n-\la\eta\xi\a_n^2}+\frac{1}
{z_2(1-2\eta\xi\a_n-\la\eta\xi\a_n^2)}\ ,\ f_0^2=0\ . 
\label{rec00}
\eea
Apart from this infinite-dimensional representation there are two
kinds of finite-dimensional ones.
The first kind is simply obtained by imposing the constraint $f_N=0$
on the parameters $z_i$, which leads to the decoupling of an
$N\times N$ block in the upper left corner of the
infinite-dimensional representation of $A$ and $B$ discussed above.
The resulting finite-dimensional representation is given in terms of
$N\times N$ matrices $A$ and $B$ with vanishing diagonal elements.

The matrices $A$ and $B$ of the second type of $N\times N$
representation take the following form 
\be
A=\left(\matrix{0&f_1&0&0&0&0&0\ldots& &\cr
\a_1\eta f_1&0&f_2&0&0&0&0\ldots& &\cr
0&\a_2\eta f_2&0&f_3&0&0&0\ldots& &\cr
0&0&\a_3\eta f_3&0&f_4&0&0\ldots& &\cr
0&0&0&\a_4\eta f_4&0&f_5&0\ldots& &\cr
\ldots& & & & \ldots& & & &\cr
\ldots& & & & \ldots& & & &\cr
\ldots& & & & \ldots& &\a_{N-2}\eta f_{N-2} &0 &f_{N-1}\cr
\ldots& & & & \ldots& &0&\a_{N-1}\eta f_{N-1} &a_{N}\cr}\right)\ ,
\label{aaa}
\ee
\be
B=\left(\matrix{0&\a_1\xi f_1&0&0&0&0&0\ldots & &\cr
f_1&0&\a_2\xi f_2&0&0&0&0\ldots & &\cr
0&f_2&0&\a_3\xi f_3&0&0&0\ldots & &\cr
0&0&f_3&0&\a_4\xi f_4&0&0\ldots & &\cr
0&0&0&f_4&0&\a_5\xi f_5&0\ldots & &\cr
\ldots& & & &\ldots& & & &\cr
\ldots& & & &\ldots& & & &\cr
\ldots& & & & \ldots& &f_{N-2}&0&\a_{N-1}\xi f_{N-1} \cr
\ldots& & & & \ldots& &0&f_{N-1}&b_N&\cr}\right)\ ,
\label{bbb}
\ee
where $\a_n$ and $f_n$ are determined by the recursion \r{rec00}.
The representation \r{aaa}, \r{bbb} exists provided that
\be
\a_N^2\eta\xi=1\ .
\ee
The variables $a_N$ and $b_N$ are obtained from the equations
\bea
b_N&=&\xi\a_N a_N\ ,\nn
\left(-\la -2\xi\eta\a_{N-1}+\eta\xi\a_{N-1}^2\right)f_{N-1}^2&=&
\xi a_N^2+\eta b_N^2 +(\la-1)a_Nb_N+\frac{1}{z_2}\ .
\eea

We are going to close this section with two cases, for which the
recurrence relations can be solved in a trivial way.
\begin{itemize}
\item{}
If $\la=-\eta\xi\neq -1$ (this case is as we will see physically
interesting) the following simplifications take place for $x_5\neq
0\neq x_6$: 
\bea
a_n&=&a=\frac{1+2\eta -\la}{z_2(1+\la)^2}\ ,\ \forall n\geq 3\ ,
a_1=0,\ a_2=\frac{1+\eta}{z_2(1+\la)}\ ,\nn
b_n&=&b=\frac{1+2\xi -\la}{z_2(1+\la)^2}\ ,\ \forall n\geq 3\ ,
b_1=0,\ b_2=\frac{1+\xi}{z_2(1+\la)}\ ,\nn
f_n^2&=&f^2=\frac{\xi a^2+\eta b^2+(\la -1)ab
+\frac{a+b+1}{z_2}}{(1+\la)^2}\ ,\ \forall n\geq 3\ ,\nn
f_1^2&=&\frac{1}{z_2},\ f_2^2=\frac{1}{z_2(1+\la)^2}\left(
1+\la+\frac{1+\xi+\eta-\la}{z_2(1+\la)}\right) \ .
\label{70}
\eea
Note that in this case there exist only 2-d ($f_2=0$), 3-d ($f_3=0$),
and infinite-dimensional representations !
If $x_5=x_6=0$, one has an infinite-dimensional representation
with vanishing diagonal elements $a_n=b_n=0\ ,\ \forall n$ with
\be
\a_n=1\ \forall n\geq 2,\quad f_1^2=\frac{1}{z_2}\ ,\
f_2^2=\frac{1}{z_2(1+\la)}\ ,\ 
f_n^2=\frac{1}{z_2(1+\la)^2}\ ,\ \forall n\geq 3\ ,
\ee
where $\eta,\xi,\la$ are given by \r{exl2}.

\item{}\underbar{$0=\eta=\xi$}

Here it is possible to choose $\tB=\tA^\dagger$, and
\bea
\a_n&=&\frac{1-\la^{n-1}}{1-\la}\ ,\
a_n=\frac{1}{z_2}\left(\frac{1-\la^{n-1}}{1-\la}\right)\ ,\quad
f_n^2=\frac{1}{z_2}\left(\frac{1-\la^{n}}{1-\la}\right)
\left(1+\frac{1}{z_2}\frac{1-\la^{n-1}}{1-\la}\right),
\label{72}
\eea
where we have assumed $x_5\neq 0\neq x_6$. A similar simplification
holds if $x_5=0$ or $x_6=0$.
Of particular physical interest is the case $\la=1$, which exhibits
additional simplifications 
\be
\a_n=n-1\ ,\qquad a_n= \frac{n-1}{z_2}\ ,\ f_n^2=
\frac{n}{z_2}+\frac{n(n-1)}{z_2^2}\ . 
\label{71}
\ee
\end{itemize}

\section{The Master Equation and the Quantum Chain Hamiltonian}

Let us consider a one-dimensional open chain with $L$ sites.
On each site $k (k = 1, 2, ... L)$ we allow for two configurations
described by means of the variable $\tau_k$, which takes the two
values $0$ and $1$. For $\tau_k = 0$ the site $k$ is empty (vacancy),
for $\tau_k = 1$ the site $k$ is occupied by a molecule $A$. 

At time $t$ the probability to find a certain configuration of
molecules and vacancies on the chain is given by the probability
distribution 
\be
P_L\left( \tau_1, \tau_2, \ldots \tau_L |t\right) \ .
\ee
If we assume that interaction between molecules is described by
two-body processes only (three-body processes are considered in
Appendix B), the time evolution of the system is given by a master
equation of the form 
\bea
\frac{\partial P_L}{\partial t}&=&-\sum_{k=1}^{L-1}\sum_{\g_k,\g_{k+1}}
\left(H_{k,k+1}\right)_{\tau_k,\tau_{k+1}}^{\g_k,\g_{k+1}}
P_L(\tau_1,\tau_2\ldots\tau_{k-1},\g_k,\g_{k+1},\tau_{k+2}\ldots\tau_L|t)\nn
&&-\sum_{\g_1}\left(h_{1}\right)_{\tau_1}^{\g_1}
P_L(\g_1,\tau_2\ldots\tau_L|t)
-\sum_{\g_L}\left(h_{L}\right)_{\tau_L}^{\g_L}
P_L(\tau_1,\tau_2\ldots\tau_{L-1},\g_L|t)\ ,
\label{mastereq}
\eea
where the boundary contributions $h_1$ and $h_L$ describe injection
(extraction) of particles with rates $\a$ and $\d$ ($\g$ and $\b$) at
sites $1$ and $L$ 
\be
h_1=\left(\matrix{\a&-\g\cr -\a&\g\cr}\right)\ ,\ 
h_L=\left(\matrix{\d&-\b\cr -\d&\b\cr}\right)\ ,
\label{h1l}
\ee
and where
\be
\left(H_{k,k+1}\right)_{\tau_k,\tau_{k+1}}^{\g_k,\g_{k+1}}
=\Bigg\lbrace{\matrix{\sum'_{\b_k,\b_{k+1}}
[{\Gamma_{k,k+1}}]_{\b_k,\b_{k+1}}^{\g_k,\g_{k+1}}\qquad
\g_j=\tau_j\ ,\ j=k,k+1\cr
-[{\Gamma_{k,k+1}}]_{\tau_k,\tau_{k+1}}^{\g_k,\g_{k+1}}\qquad {\rm else}\cr}}
\label{hk}
\ee
Here $[\Gamma_{k,k+1}]^{\gamma_k , \gamma_{k + 1}}_{ \tau_k ,
\tau_{k + 1}} $ represents the probability per unit time that the
configuration $(\gamma_k , \gamma_{k + 1})$ on neighbouring sites $k$
and $k+1$ changes into the configuration $(\tau_k , \tau_{k + 1})$ and
$\sum'$ denotes the sum where the term
$(\g_k,\g_{k+1})=(\beta_k,\beta_{k+1})$ is excluded.
The following processes are included in the master equation (with
$0$ a vacancy $(\tau = 0)$ and $A$ a molecule $(\tau = 1 )$)

\be
\begin{array}{lcr}
{\mbox {Diffusion to the right:}} & A + 0 \to 0 + A & ({\mbox{rate}}\;
\Gamma^{10}_{01}) \nonumber \\
{\mbox {Diffusion to the left:}} & 0 + A \to A + 0 & (\Gamma^{01}_{10})
\nonumber \\
{\mbox {Coagulation at the right:}} & A + A \to 0 + A & (\Gamma^{11}_{01})
\nonumber \\
{\mbox {Coagulation at the left:}} & A + A \to A + 0 & (\Gamma^{11}_{10})
\nonumber \\
{\mbox {Decoagulation at the right:}} & A + 0 \to A + A & (\Gamma^{10}_{11})
\nonumber \\
{\mbox {Decoagulation at the left:}} & 0 + A \to A + A & (\Gamma^{01}_{11})
\nonumber \\
{\mbox {Birth at the right:}} & 0 + 0 \to 0 + A & (\Gamma^{00}_{01}) \nonumber
\\
{\mbox {Birth at the left:}} & 0 + 0 \to A + 0 & (\Gamma^{00}_{10}) \nonumber
\\
{\mbox {Death at the right:}} & 0 + A \to 0 + 0 & (\Gamma^{01}_{00}) \nonumber
\\
{\mbox {Death at the left:}} & A + 0 \to 0 + 0 & (\Gamma^{10}_{00}) \nonumber
\\
{\mbox {Pair-annihilation:}}	      & A + A \to 0 + 0 & (\Gamma^{11}_{00})
\nonumber \\
{\mbox {Pair-creation:}}	      & 0 + 0 \to A + A & (\Gamma^{00}_{11})
\label{rates}
\end{array}
\ee
Reaction-diffusion models of the type described above can be mapped to
quantum spin chains in the following way \cite{vlad1}:
a basis of the quantum-mechanical Hilbert space ${\cal H}$ (isomorphic
to the tensor product $\otimes_{n=1}^L C^2$) is defined as
\be
|\{\tau\}\rangle = |\tau_1\ldots\tau_L\rangle\ ,
\ee
and the inner product is taken as
\be
\langle \{\tau\}|\{\tau'\}\rangle = \prod_{j=1}^L\d_{\tau_j,\tau_j'}.
\ee
This induces a map of the probability distribution $P_L$ to a state in
${\cal H}$
\be
|P\rangle = \sum_{\{\tau\}}P_L(\tau_1\ldots\tau_L|t) |\{\tau\}\rangle\ ,
\label{qmstate}
\ee
and the master equation \r{mastereq} then implies an imaginary-time
Schr\"odinger equation
\be
\frac{\partial|P\rangle}{\partial t} = -{\hat H} |P\rangle \ .
\label{masterqm}
\ee
Here ${\hat H}$ is a quantum hamiltonian defined in terms of a basis
$E_k^{\a\b}$ (which can be represented as $2\times 2$ matrices with
entries ${E^{\a\b}}_{\g\d} = \d_{\a\g}\d_{\b\d}$) of quantum operators
on the $k$'th site of the lattice {\sl via}
\be
{\hat H} = \sum_{k=1}^{L-1}
(H_{k,k+1})^{\a\b}_{\g\d}E_k^{\g\a}E_{k+1}^{\d\b} +
(h_1)^{\a}_\g E_1^{\g\a}+(h_L)^{\a}_\g E_L^{\g\a}\ ,
\label{ham}
\ee
where $h_1$, $h_L$ and $H_{k,k+1}$ are defined in \r{hk}.
Note that in general the hamiltonian ${\hat H}$ will be non-hermitian.
It is easy to see that
\be
\langle 0| {\hat H} = 0\ ,
\label{bra}
\ee
where $\langle 0 \mid$ is given by
\be
\langle 0 \mid = \sum_{\{ \tau \}} \langle \{ \tau \} \mid \; = \;
\left\langle {1 \choose 1} \otimes  \right.
{1 \choose 1} \otimes \ldots \otimes {1 \choose 1}  \mid
\ee
Using \r{masterqm} it follows from \r{bra} that $\langle 0|$ is a left
``stationary'' state. Assuming that this is the unique left ``stationary''
state and given a unique right stationary state 
\be
|0\rangle = \sum_{\{\tau\}}P_s(\{\tau\})|\{\tau\}\rangle
\ee
the average of the observable $X \left( \tau_1 , \ldots \tau_L
\right) = X (\{ \tau \})$ is defined as
\begin{eqnarray}
\langle X\rangle  & = & \sum_{\{ \tau \}} X (\{ \tau \} ) P_s (
\{ \tau \}) \nonumber \\
& = & \langle 0 \mid X \mid 0 \rangle \ .
\end{eqnarray}

An example, with which we will be concerned in most of this paper is the
case of partially asymmetric diffusion, which corresponds to
the choice of rates
\be
[{\G_{k,k+1}}]^{10}_{01}=p\ ,\  [{\G_{k,k+1}}]^{01}_{10}=q
\ee
(all other rates are taken to be zero) the quantum hamiltonian ${\hat H}$
obtained by the above mapping is related to an XXZ spin chain by a
similarity transformation
\be
H_{XXZ}=U^{}{\hat H}U^{-1}\ ,\ 
U=\prod_{j=1}^L\left(E_j^{00}+E_j^{11}\La{\cal Q}^{j-1}\right)
=\prod_{j=1}^L\left(\matrix{1&0\cr 0& \La{\cal Q}^{j-1}\cr}\right)
\label{sim} 
\ee
where ${\cal Q}=\sqrt{\frac{q}{p}}$, $\La$ is a free parameter and
\begin{eqnarray}\label{A5d}
\frac{1}{\sqrt{pq}}H_{XXZ}&=&- \frac{1}{2}
\sum_{j=1}^{L-1}\biggl[2(\sigma^+_j \sigma^-_{j+1}+\sigma^-_j
\sigma^+_{j+1}) +\frac{1}{2}({\cal Q}+{\cal Q}^{-1}) \sigma^z_j
\sigma^z_{j+1}\nn
&&\qquad\qquad +\frac{{\cal Q}-{\cal Q}^{-1}}{2}
(\sigma_{j+1}^z-\sigma_{j}^z) -\frac{1}{2}({\cal Q}+{\cal Q}^{-1}) 
\biggr]+B_1+B_L \nn
B_1&=&\sigma_1^z\frac{\a-\g}{2\sqrt{pq}}-\sigma_1^-\frac{\La\a}{\sqrt{pq}}
-\sigma_1^+\frac{\g}{\La\sqrt{pq}}+\frac{\a+\g}{2\sqrt{pq}}\ ,\nn
B_L&=&\sigma_L^z\frac{\d-\b}{2\sqrt{pq}}-\sigma_L^-\frac{\La\d}{\sqrt{pq}}
{\cal Q}^{L-1}-\sigma_L^+\frac{\b}{\La\sqrt{pq}}{\cal Q}^{1-L}
+\frac{\b+\d}{2\sqrt{pq}}\ .
\label{hxxz}
\end{eqnarray}

This is the $U_q(SU(2))$ invariant quantum spin chain \cite{ps} with
added boundary terms $B_1$ and $B_L$. Notice that the boundary terms
contain nondiagonal contributions ($\sigma_1^\pm$, $\sigma_L^\pm$) with
$L$-dependent coefficients. In the absence of the boundary terms the
spectrum of the hamiltonian is massive. As is shown below the boundary
terms with generate phase transitions with massless phases.
Although the hamiltonian \r{hxxz} can be shown to be integrable
\cite{deV}, the Bethe Ansatz so far has not been constructed due to
the lack of a reference state.

We note that the similarity transformation \r{sim} does not change
averages of observables 
\be
\langle 0|X|0\rangle = \langle 0|U^{-1}UXU^{-1}U|0\rangle =\ 
_U\!\langle 0|X_U|0\rangle _U\ .
\ee
Thus zero-temperature equal time correlation functions of the XXZ
quantum spin chain and (stationary-state) averages of the partially
asymmetric diffusion model are related in the following way
\bea
\langle \sigma^z_j\rangle&=&_U\!\langle 0|\sigma^z_j|0\rangle_U=
2\langle \tau_j\rangle -1\ ,\nn
\langle \sigma^z_j\sigma^z_k\rangle&=& _U\!\langle 0|\sigma^z_j
\sigma^z_k|0\rangle_U= 4\langle \tau_j\tau_k\rangle +1-2\langle\tau_j\rangle
-2\langle\tau_k\rangle\ ,\nn
\langle \sigma^z_j\sigma^z_k\rangle_{conn}&=&
4\langle \tau_j\tau_k\rangle_{conn}\ ,
\label{trans}
\eea
where $_{conn}$ denotes connected correlation functions.
This means that all results concerning averages in the partially
asymmetric diffusion model obtained in this paper can be immediately
applied to the case of the XXZ chain described above.

\section{The DEHP Ansatz}

In a remarkable paper \cite{dehp} it was shown that for the case of
asymmetric diffusion the problem of determining the probability
distribution $P_L(\tau_1,\ldots \tau_L)$ for a {\sl stationary state}
can be formulated in a completely algebraic framework.
We now briefly review the relevant results. All rates except
$[\G_{k,k+1}]^{10}_{01}=p$ and $[\G_{k,k+1}]^{01}_{10}=q$ ($k=1,\ldots
L-1$) are taken to be zero, and the boundary conditions are chosen
according to \r{h1l}: particles are injected at sites $1$ and $L$ with
rates $\a$ and $\d$ and extracted with rates $\g$ and $\b$ respectively.
The algebraization of the problem of determining the {\sl
unnormalized} probability distribution $P_L(\tau_1\ldots\tau_L)$ of a
stationary state is performed in two steps: one first makes
an Ansatz $P_L(\tau_1,\ldots \tau_L)$ in the form of a matrix-product
state \cite{dehp,matprod} 

\begin{eqnarray} \label{B1}
P_L(\tau_1,...,\tau_L)&=&\langle W| \prod_{i=1}^L (\tau_iD+(1-\tau_i)E)|V
\rangle\ .
\end{eqnarray}
Here $D$ and $E$ are in general infinite-dimensional matrices and $\w$
and $\v$ are vectors connected with the boundary conditions.
The normalization factor is obviously given by
\be
Z_L= \w C^L\v\ ,\qquad C=D+E\ .
\ee
In the second step the following sufficient conditions for $P_L$ to be
a stationary solution of the master equation are imposed
\bea
\sum_{\g_k,\g_{k+1}}
\left(H_{k,k+1}\right)_{\tau_k,\tau_{k+1}}^{\g_k,\g_{k+1}}&&
P_L(\tau_1,\tau_2\ldots\tau_{k-1},\g_k,\g_{k+1},\tau_{k+2}\ldots\tau_L) \nn
=&& x_{\tau_k} P_{L-1}(\tau_1\ldots\tau_{k-1}\tau_{k+1}\ldots \tau_L)
 -x_{\tau_{k+1}} P_{L-1}(\tau_1\ldots\tau_{k}\tau_{k+2}\ldots
\tau_L)\ ,
\label{algstat}\\
&&\sum_{\g_1}^{}(h_{1})_{\tau_1}^{\g_1}P_L(\g_1\tau_2\ldots\tau_L)=
-x_{\tau_1}P_{L-1}(\tau_2\ldots\tau_L)\nn
&&\sum_{\g_L}(h_{L})_{\tau_{L}}^{\g_L}P_L(\tau_1\ldots\tau_{L-1}\g_{L} )=
x_{\tau_{L}}P_{L-1}(\tau_1\ldots\tau_{L-1})\ .
\label{stat1}
\eea
Inserting \r{B1} into \r{algstat} , \r{stat1} leads to algebraic
relations between the matrices $D$ and $E$ and leads to conditions for
the action of $D$ and $E$ on $\v$ and $\w$ \cite{dehp} (one finds that
$x_0=-x_1$ and then sets $x_1=1$)
\bea
pDE-qED&=&D+E\nn
(\b D-\d E)\v&=&\v\nn
\w (\a E-\g D)&=& \w \nn
\langle W|V\rangle&\neq& 0\ .
\label{de}
\eea
It is easy to see that for $\a\b=\g\d$ no representations of
\r{de} exist. This can be seen by considering the inner product
\be
\w \left(\a E-\g D\right)\v \ .
\label{abcd}
\ee
Evaluating \r{abcd} once by acting to the left and once by acting to
the right using that $\a E-\g D = -\frac{\g}{\b}(\b D-\d E)$ 
(which holds because $\a\b=\g\d$) we obtain
\be
\langle W|V\rangle = -\frac{\g}{\b}\langle W|V\rangle\ ,
\ee
which has the only solution $\langle W|V\rangle =0$.
This means that from now on we can constrain ourselves without loss
of generality to the case $\a\b\neq \g\d$\footnote{Actually there
is one exception: if $p=q$ and $\a\b=\g\d$ there exists a trivial
one-dimensional representation ($D$ and $E$ are numbers) with $x_0=0$,
$D=\frac{\d}{\b}E$.}. Moreover one can show that all vacuum
expectation values
\be
\w D^{r_1}E^{r_2}\ldots D^{r_n}\v
\ee
are determined by \r{de} if the following inequalities are satisfied
\be
p^k\a\b-q^k\g\d\neq 0\ ,\quad k=0,1,2,\ldots
\label{56}
\ee
The proof uses the construction of \cite{sandow} and one can show that if
\r{56} is satisfied a representation exists even if its dimension is not
the one of the smallest representation.
In order to obtain $P_L$ it is now necessary to find matrices $D$ and
$E$ together with vectors $\w$ and $\v$ obeying \r{de}.

For later use we note that physical quantities like the
current $J$, density profile $\langle\tau_j\rangle$ and two-point
function $\langle\tau_j\tau_k\rangle$ can be evaluated in the
following way 
\cite{dehp} 
\bea
J&=& \frac{\w C^{L-1}\v}{\w C^{L}\v}\ ,\nn
\langle \tau_j\rangle&=& \frac{\w C^{j-1}DC^{L-j}\v}{\w C^{L}\v}\ ,\nn
\langle \tau_j\tau_k\rangle&=& \frac{\w C^{j-1}DC^{k-j-1}DC^{L-k}\v} 
{\w C^{L}\v}\ ,
\label{jttt}
\eea
where $C=D+E$.

It is possible to determine certain matrix elements of representations
of the six-parametric algebra \r{de} directly. This was done by Sandow
\cite{sandow} who then was able to determine both the current $J$ (the
computations of which involves only the matrix $C$) in the infinite
volume limit $L\rightarrow\infty$ and, remarkably, the phase diagram
of the system. 
In order to compute the density profile and the two-point function a
much more detailed understanding of the representations is needed.
In order to study representation theory of \r{de} we introduce
two operators $A$ and $B$, which act trivially on $\w$ and $\v$
respectively 
\bea
A&=&\b D-\d E-1\ ,\ B=\a E - \g D -1 \ ,\nn
\w B&=&0=A\v\ ,
\eea
If $A$ and $B$ are known, one can get $D$ and $E$ since $\a\b\neq\g\d$.
$A$ and $B$ are seen to obey the quadratic algebra \r{ab} discussed in
section II with
\bea
x_1&=&(p-q)\a\g\ ,\quad x_2=p\a\b-q\g\d\ ,\quad x_3=p\g\d-q\a\b\
,\quad x_4=(p-q)\b\d\ ,\nn
x_5&=&(\a+\g)(\a\b-\g\d)-(p-q)[\g(\a+\d)+\a(\b+\g)]\ ,\nn
x_6&=&(\b+\d)(\a\b-\g\d)-(p-q)[\b(\a+\d)+\d(\b+\g)]\ ,\nn
x_7&=&(\a+\b+\g+\d)(\a\b-\g\d)-(p-q)(\a+\d)(\b+\g)\ .
\label{x}
\eea

Note that the seven parameters $x_i$ are not independent, since they
depend only on six variables $p,q,\a,\b,\g,\d$. Conversely the algebra
\r{ab} can be brought to the form \r{de} if
\be
(x_2+x_3)^2-4x_1x_4\neq 0\ .
\label{57}
\ee
One can easily verify that the inequalities \r{57} and \r{vier} correspond
to those of \r{56} for $k=0,1,2,3$.

We now observe that in the case $q=0$ of completely asymmetric diffusion
we have $\la=-\eta\xi$ and the matrices $\tA$ and $\tB$ (which we
recall were defined as $\tA=\frac{x_5}{x_7}A$ and
$\tB=\frac{x_6}{x_7}B$) have the simple forms given by \r{70}.
The origin of this simplification can be traced back to the
representation theory of quantum groups which also simplifies
drastically in the crystal basis $q=0$ \cite{kash}. Another case for
which $A$ and $B$ have a simple form is $p=q$ (symmetric diffusion).
Here we have $ \xi =\eta=0$,$\la = 1$, and the representation is given
by \r{71}. Finally, if $\a =\b =p-q$, $\g =\d=0$, we can define
\be
a=\sqrt{\frac{p}{p-q}} A\ ,\ a^\dagger=\sqrt{\frac{p}{p-q}} B
\label{qos}
\ee
and the algebra \r{ab} is rewritten as a Q-oscillator
algebra\cite{kul,mac,bied}
\be
aa^\dagger - Q a^\dagger a =1\ ,
\ee
where $Q=\frac{q}{p}$. The vectors $\v$ and $\w$ turn into usual
Fock-vacua $\langle 0|$ and $|0\rangle$ defined by
$a|0\rangle=\langle 0|a^\dagger=0$. Some observations about this case
can be found in Appendix C.

As noticed before in section III, in the absence of boundary terms the
hamiltonian \r{hxxz} is $U_q(SU(2))$ invariant. The ground state of
this hamiltonian is $L+1$ times degenerate (recall that $L$ is the
length of the lattice) corresponding to a $L+1$-dimensional
representation of the algebra. We demonstrate in Appendix D that this
representation can also be found through the DEHP-Ansatz.

\section{Further Applications of the DEHP Ansatz}

It is an interesting question, to what extent the Ansatz (\ref{B1})
can be used to describe more general reaction-diffusion models of the
type (\ref{rates}). Inserting \r{B1} into the master equation \r{stat1}
for a general reaction-diffusion process defined {\sl via}
\r{h1l}-\r{rates} (note that we take all rates constant throughout the
bulk, {\sl i.e.}, $[\G_{k,k+1}]^{\a\b}_{\g\d}=\G^{\a\b}_{\g\d}$),
we obtain the algebra
\bea
&&{\cal H}\pmatrix{E^2\cr ED\cr DE\cr D^2\cr}=\pmatrix{0\cr
x_0D-x_1E\cr -x_0D+x_1E\cr 0\cr}\ , \nn
&&{\cal H}\!=\!\left(\matrix{
\G^{00}_{01}\!+\!\G^{00}_{10}\!+\!\G^{00}_{11} &\!\!  -\G^{01}_{00}&\!\! 
-\G^{10}_{00}&\!\!  -\G^{11}_{00}\cr 
-\G^{00}_{01}&\!\!  \G^{01}_{00}\!+\!\G^{01}_{10}\!+\!\G^{01}_{11}
&\!\!  -\G^{10}_{01}&\!\!  -\G^{11}_{01}\cr
-\G^{00}_{10}&\!\!  -\G^{01}_{10}&\!\!
\G^{10}_{00}\!+\!\G^{10}_{01}\!+\!\G^{10}_{11} &\!\!  -\G^{11}_{10}\cr 
-\G^{00}_{11}&\!\!  -\G^{01}_{11}&\!\!  -\G^{10}_{11}&\!\! 
\G^{11}_{00}\!+\!\G^{11}_{01}\!+\!\G^{11}_{10} \cr}\right),
\label{genalg}
\eea
whereas the boundary conditions \r{stat1} impose the following
conditions on the vectors $\v$ and $\w$
\bea
&&\sum_{\g_1}(h_{1})_{\tau_1}^{\g_1}\w [\g_1D+(1-\g_1)E]= -x_{\tau_1}\w\nn
&&\sum_{\g_L}(h_{L})_{\tau_L}^{\g_L}
[\g_LD+(1-\g_L)E]\v=x_{\tau_L}\v\ . 
\label{bcs}
\eea
Here $h_1$ and $h_L$ are given by \r{h1l} and thus \r{bcs} are
independent from \r{genalg}. From \r{bcs} it follows that $x_0=-x_1$,
so that we can choose $x_0=-1$, $x_1=1$ by fixing the overall
normalization in \r{stat1}.

It is easily seen that only three
equations of (\ref{genalg}) are linearly independent. These can be
cast in the form
\bea
\kappa_1 DE + \kappa_2 ED &=& D+E\ ,\label{diffalg1}\\
\kappa_3D^2 &=&\kappa_4 DE + \kappa_5 ED\ ,\label{diffalg2}\\
\kappa_6E^2 &=&\kappa_7 DE + \kappa_8 ED\ ,\label{diffalg3}
\eea
where $\kappa_j$ are given in terms of the rates $\G^{\a\b}_{\g\d}$.
These equations can be viewed in the following way:
(\ref{diffalg1}) is the basic requirement as it cannot be eliminated by
adjusting the rates, whereas (\ref{diffalg2}) and (\ref{diffalg3}) are
additional relations in the algebra which are absent in the simplest
case where $\kappa_3\ldots\kappa_8$ are chosen to be zero (by adjusting
the rates). This simplest case corresponds to partially asymmetric
diffusion and will be studied in detail in what follows. The important
point is that the set of representations of
(\ref{diffalg1})-(\ref{diffalg3}) is a subset of all representations
of (\ref{diffalg1}) for arbitrary $\kappa_1$ and $\kappa_2$.
This means that in all cases physical quantities can be determined by
using a representation of (\ref{diffalg1}) only, and then impose the
further relations on the matrix elements of $D$ and $E$ entering the
computation. The existence of solutions of the complete system
\r{bcs}, \r{diffalg1}-\r{diffalg3} is established for the simple case
of one-dimensional representations of in Appendix E.
The question of existence of a two-dimensional representation is
still open.

\section{Asymmetric Diffusion: Known Results}

Before we turn to the derivation of our results for current, density
profile and correlation functions of the partially asymmetric
exclusion model we give a short review of some important previously
known exact results. So far exact results have mainly been derived for
the case of completely asymmetric diffusion with injection of
particles at one boundary, and extraction at the other. In our
notation this corresponds to the choice $0=q=\g=\d$, $p=1$. 
This corresponds to the infinite-dimensional representation given by
\r{72} with $0=\la=\eta=\xi$. The phase diagram for this case is of the
form given in Fig.1 \cite{ddm,dehp,sd}. Note that in order to make the
connection to the
partially asymmetric case easier we have plotted the phases as
functions of $\kappa_+(\a)=-1+\frac{1}{\a}$ and
$\kappa_+(\b)=-1+\frac{1}{\b}$ instead of $\a$ and $\b$. There are
three main phases: a high-density phase $A$, a low-density phase $B$,
and a maximal-current phase $C$. Phases $A$ and $B$ are further
subdivided into $A_I$, $A_{II}$ and $B_I$, $B_{II}$ respectively
\cite{sd} (see below; note that we have 
changed notations by switching $A$ and $B$ as compared to \cite{sd} in
order to comply with the notation of \cite{sandow}). Phases $A$ and $B$
are separated by a line which is called ``coexistence line''.

\begin{figure}[htbtv]
\begin{center}
\setlength{\epsfysize}{100mm}
\setlength{\epsfxsize}{100mm}
\leavevmode
\epsfbox{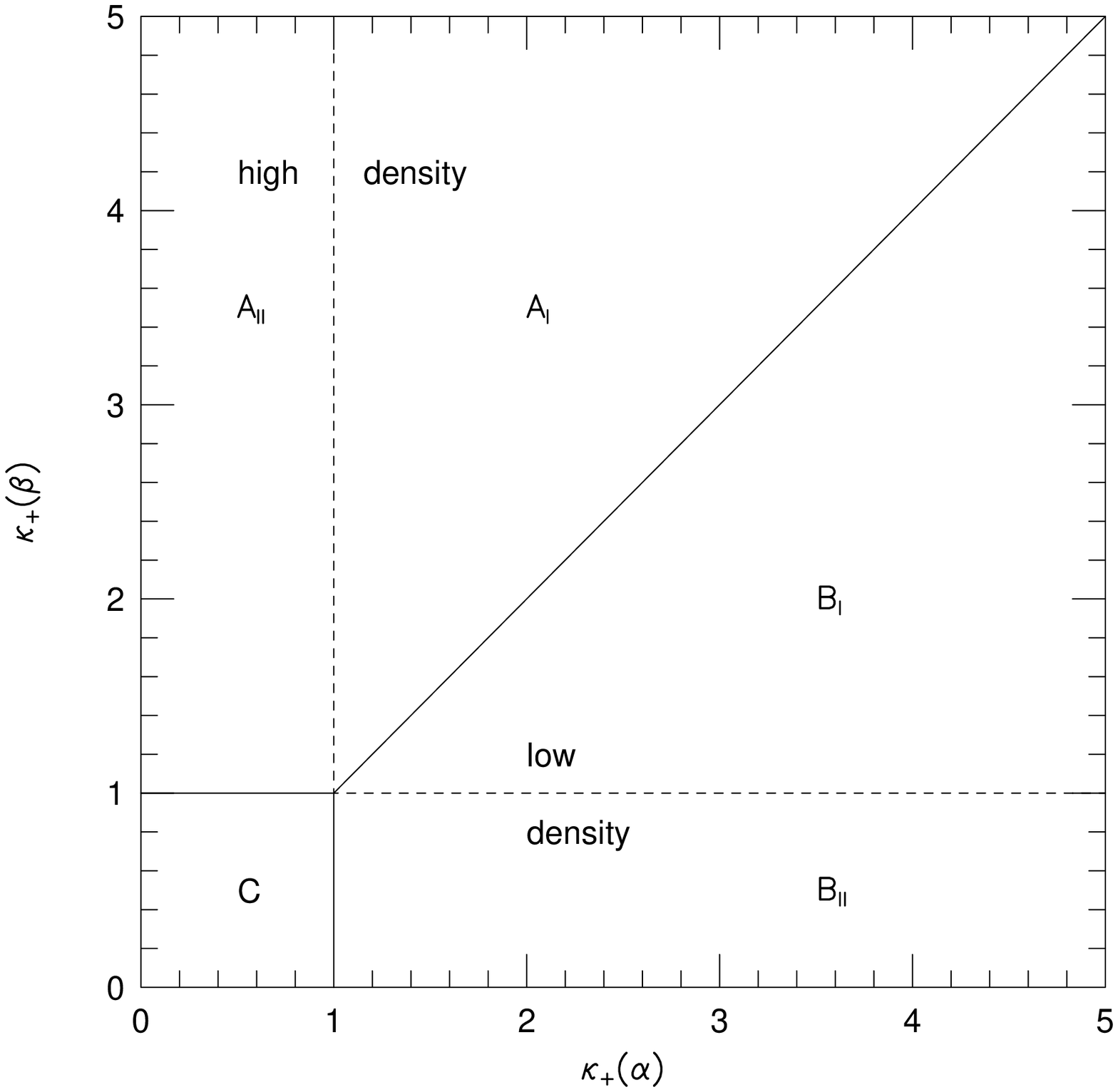}
\end{center}
\caption{\label{pd2}}
{\sl Phase diagram of the completely asymmetric exclusion model.}
\end{figure}

The currents in the three phases are given by ($ L\gg 1$)
\bea
{\rm Phase\   A :}\quad J&=&\b(1-\b)  \ ,\nn
{\rm Phase\   B :}\quad J&=&\a(1-\a)  \ ,\nn
{\rm Phase\   C :}\quad J&=& \frac{1}{4} \ .
\eea
The density profile in the center of the chain ($j\sim\frac{L}{2}$, 
$L\gg 1$) is of the form 
\bea
{\rm Phase\   A\  (high\ density)\ :}\quad \langle\tau_j\rangle
&=&\frac{\kappa_+(\b)}{1+\kappa_+(\b)}  \ ,\nn 
{\rm Phase\   B\  (low\ density)\ :}\quad \langle\tau_j\rangle
&=&\frac{1}{1+\kappa_+(\a)}  \ ,\nn 
{\rm coexistence\ line :}\quad \langle\tau_j\rangle
&=& \a +(1-2\a)
\left(\frac{j}{L}\right)\ ,\nn 
{\rm Phase\   C :}\quad \langle\tau_j\rangle &=& \frac{1}{2} \ .
\label{leading}
\eea
The subdivision of phase $A$ into $A_I$ and $A_{II}$ (and similarly
$B$ in to $B_I$ and $B_{II}$) was proposed in \cite{sd} and is based
on an analysis of the behaviour of the density profile near the ends
of the chain, which for $L\gg j\gg 1$ is of the form \cite{dehp}

\bea
{\rm Phase\   A_I :}\quad \langle\tau_{j}\rangle 
&=&\frac{\kappa_+(\b)}{1+\kappa_+(\b)}-(1-2\a)\left[\frac{\kappa_+(\b)}
{\kappa_+(\a)}\left(\frac{1+\kappa_+(\a)}{1+\kappa_+(\b)}\right)^2
\right]^{j}\ ,\nn  
{\rm Phase\   A_{II} :}\quad \langle\tau_{j}\rangle 
&=&\frac{\kappa_+(\b)}{1+\kappa_+(\b)}-
\frac{4^{j-1}[\b(1-\b)]^{j}}{\sqrt{\pi}j^\frac{3}{2}}\left(\frac{1}
{(1-2\a)^2}-\frac{1}{(1-2\b)^2}\right)\ ,\nn
{\rm Phase\   B_I :}\quad \langle\tau_{L-j}\rangle 
&=&\frac{1}{1+\kappa_+(\a)}+(1-2\b)\left[\frac{\kappa_+(\a)}{\kappa_+(\b)}
\left(\frac{1+\kappa_+(\b)}{1+\kappa_+(\a)}\right)^2\right]^{j+1}\ ,\nn 
{\rm Phase\   B_{II} :}\quad \langle\tau_{L-j}\rangle 
&=&\frac{1}{1+\kappa_+(\a)}+\frac{4^j[\a(1-\a)]^{j+1}}
{\sqrt{\pi}j^\frac{3}{2}}\left(\frac{1}{(1-2\b)^2}-\frac{1}{(1-2\a)^2}
\right)\ ,\nn
{\rm Phase\   C :}\quad \langle\tau_{L-j}\rangle 
&=&\frac{1}{2}-(1-\d_{\b,\frac{1}{2}})\frac{1}{2\sqrt{\pi}\ j^\frac{1}{2}}\ .
\label{asdens}
\eea
Note that if $\b=\frac{1}{2}$ in phase C there is no $j$-dependent
correction terms in the density.
The mixed notation in terms of $\kappa_+(\a)$, $\kappa_+(\b)$ and
$\a$, $\b$ has been chosen deliberately and is based on universal
behaviour in the partially asymmetric case (see below).

Correlation functions for the completely asymmetric case ($0=q=\g=\d$,
$p=1$) and $\a=1=\b$ were obtained in \cite{de}. This corresponds to
the point $0=\kappa_+(\a)=\kappa_+(\b)$ in the phase diagram (this is
the Q-oscillator representation \r{qos} with $Q=0$). In the
thermodynamic limit $L\rightarrow\infty$, $k_1\gg 1$, $k_2\gg 1$,
$k_j$ fixed, the connected two-point function was found to exhibit
an algebraic decay
\be
\langle\tau_{k_1}\tau_{k_1}\rangle -\langle\tau_{k_1}\rangle
\langle\tau_{k_2}\rangle = -\frac{1}{4(k_1k_2)^\frac{1}{2}}\left[
1-\left(1-\frac{k_2}{k_1}\right)^\frac{1}{2}\right]\ .
\ee
Finally, in \cite{ss} the density profile in the point $p=q$
($\a,\b,\g,\d$ arbitrary) was computed. This again corresponds to a
simple representation of the algebra (see \r{71}).

Much less is known about the partially asymmetric diffusion process.
The current $J$ in the large-$L$ limit was determined in
\cite{sandow}. In analogy with the case of completely asymmetric
diffusion discussed above, a phase diagram with only three different
phases was proposed on the basis of the form of the current. The relevant
variables for determining the phases are $\kappa_+(\a,\g)$ and
$\kappa_+(\b,\d)$, where
\begin{equation} \label{C2}
\kappa_{+}(x,y)=\frac{1}{2 x}[-x+y+p-q +\sqrt{(-x+y+p-q)^2+4xy }\;]
\;\;.\end{equation}
Note that for $0=y=q,\ p=1$ this definition reduces to the one for
$\kappa_+(\a)$ (see above).
In terms of these variables the current phase-diagram 
exhibits the following three phases \cite{sandow}\vspace*{.5cm}

{\em \underbar{Phase A}: $\kappa_+(\beta,\delta)>\kappa_+(\alpha,\gamma)$ ,
$\kappa_+(\beta,\delta)>1$.}
In the limit $L\rightarrow\infty$ the current $J$ is
\begin{equation} \label{C9}
J = \frac{1}{2(p-q)}\;\{ \;(\beta-\delta)(p-q)-(\beta+\delta)^2+
(\beta+\delta)
\sqrt{(\beta-\delta-p+q)^2+4 \beta \delta}\; \}
\;\;.\end{equation}\vspace*{.5cm}

{\em \underbar{Phase B}:
$\kappa_+(\alpha,\gamma)>\kappa_+(\beta,\delta)$ ,
$\kappa_+(\alpha,\gamma)>1$.} \begin{equation} \label{C10}
J = \frac{1}{2(p-q)}\;\{ \;(\alpha-\gamma)(p-q)-(\alpha+\gamma)^2+
(\alpha+\gamma)
\sqrt{(\alpha-\gamma-p+q)^2+4 \alpha \gamma}\; \}
\;\;.\end{equation}\vspace*{.5cm}

{\em \underbar{Phase C}: $\kappa_+(\beta,\delta)<1$ ,
$\kappa_+(\alpha,\gamma)<1$.}

\begin{equation} \label{C11}
J = \frac{p-q}{4}
\;\;.
\end{equation}
These results (which are the same as the corresponding mean-field
results derived in Appendix D) are summarized in the phase-diagram
shown in Fig. $2$ 
\cite{sandow}. 
\begin{figure}[htbtv]
\begin{center}
\setlength{\epsfysize}{100mm}
\setlength{\epsfxsize}{100mm}
\leavevmode
\epsfbox{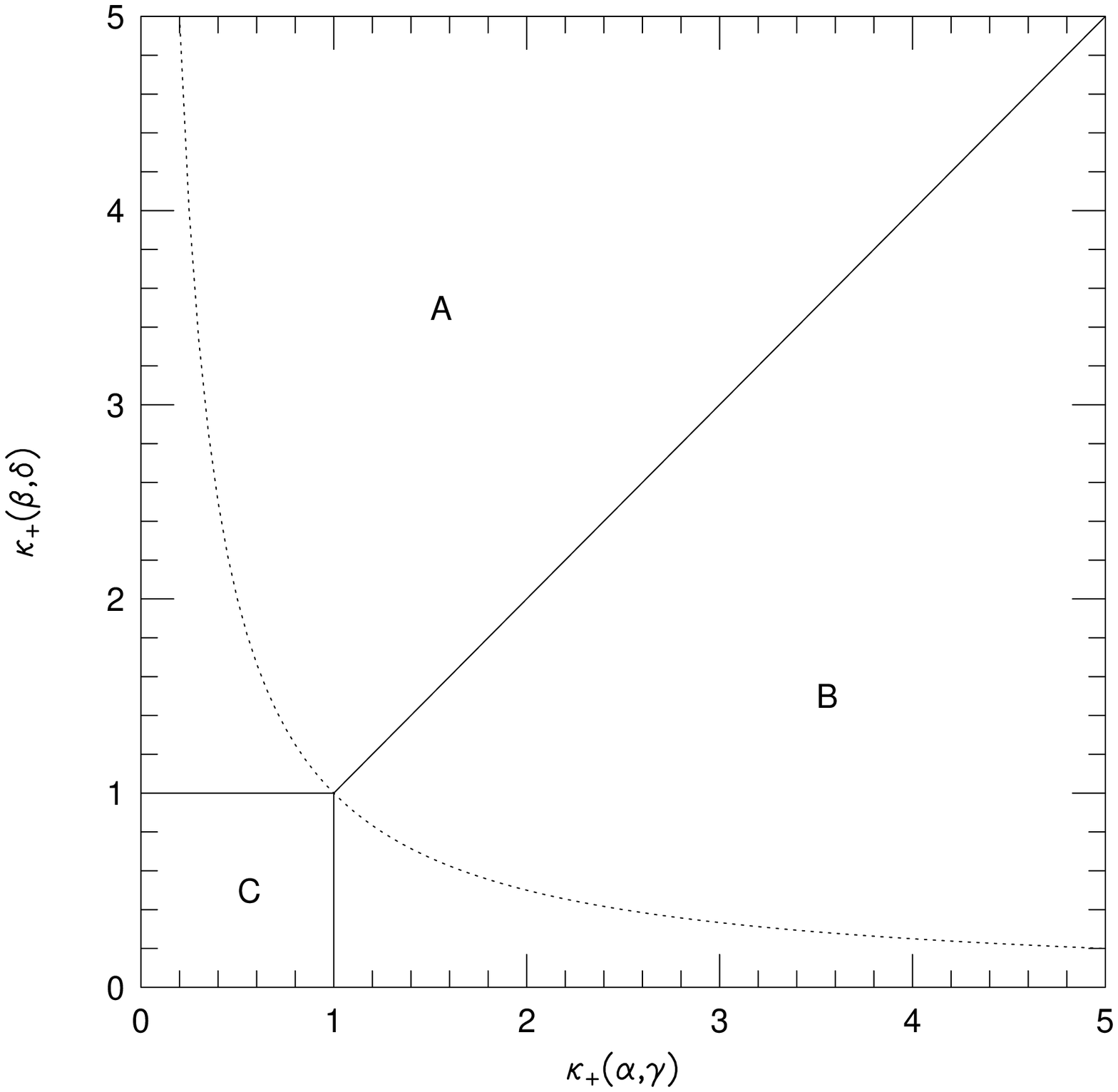}
\end{center}
\caption{\label{pd}}
{\sl Phase diagram for the current on a large lattice for $p>q$. The
regions above the dotted line in phases $A$ and $B$ are accessible by
finite-dimensional representations (see below).}
\end{figure}
We remark that in order to evaluate the current it is not necessary to
first obtain a complete representation of \r{de}, only certain matrix
elements are required \cite{sandow}. As we are interested in general
correlators we now turn to a detailed study of representations of
\r{de}. 

\section{Finite-Dimensional Representations of the Quadratic Algebra
and the Phase Diagram}

As we have seen in the last section, the calculation of the density
profile in the fully asymmetric case was done choosing the
parameters of the problem such that one obtains an
infinite-dimensional representation of a very simple form. We would
like to do the calculation of the density and the correlation functions
(which up to now are only known for $\a=\b=p=1$, $\g=\d=q=0$, which
corresponds to $0=\kappa_+(\a,\g)=\kappa_+(\b,\d)$) in a large region
of the parameter space. In order to do so, we will use the
finite-dimensional representations of the algebra. What kind of
results can one expect ? An inspection of the formulas \r{jttt}
suggests that if one writes $C=\exp(-I\!\!H\!\!I)$, $I\!\!H\!\!I$ plays
the role of a space-evolution operator although $\v$ and $\w$ are not
eigenvectors of $I\!\!H\!\!I$. Two scenarios are possible. If $C$ is
diagonizable, one expects an exponential decay of the density and
correlation functions. If $C$ is not diagonizable one can anticipate
an algebraic behaviour. From the mean-field analysis one expects an
algebraic behaviour on the coexistence line and in the domain
$\kappa_+(\a,\g) \leq 1$ and $\kappa_+(\b,\d) \leq 1$. As we will see,
the finite-dimensional representations will only access the coexistence
line. Using finite-dimensional representations makes the calculation
of the correlation functions relatively simple. However there is a
price to pay, namely one has to solve the constraint equation which
has a rather complicated expression in terms of the parameters
$\a,\b,\g,\d,p,q$.

Let us first investigate the question which regions in the phase
diagram given in Fig.2 are accessible by finite dimensional
representations. 
\begin {itemize}
\item[a)]{} One-dimensional representation

It exists whenever the constraint $x_7=0$ is
fulfilled. In terms of the variables $\kappa_+$ the constraint reads
\be
\kappa_+(\b,\d)=\frac{1}{\kappa_+(\a,\g)}\ .
\ee
It is completely straightforward to evaluate the current and the
density profile in this case. We find
\bea
J&=&\frac{\a\b-\g\d}{\a+\b+\g+\d}\ ,\nn
\langle\tau_j\rangle &=&\frac{\a+\d}{\a+\b+\g+\d}\equiv
\frac{1}{1+\kappa_+(\a,\g)}=\frac{\kappa_+(\b,\d)}{1+\kappa_+(\b,\d)}\ .
\eea
The second to last equality is established after some cumbersome
computations using the constraint $x_7=0$.

\item[b)]{} Two-dimensional representation

Let us consider the 2-d representation in detail. The constraint
$f_2=0$ is expressed in terms of the $x_i$'s as
\be
(x_2-x_3)[x_7(x_2^2-x_1x_4)^2+x_2(x_2x_5-x_1x_6)(x_2x_6-x_4x_5)]=0\ .
\label{constr2}
\ee
Taking $x_2\neq x_3$ ($x_2=x_3$ corresponds to the unphysical
situation $p=-q$) we can cast \r{constr2} in the form of a quadratic
equation for $\g$ 
\bea
0&=&c\g^2+b\g+a\ ,\nn
a&=&\b p^2 (\a^2(\b p +\d q +pq)+\a q(\b p-\d p-p^2+2\d q+pq)-\d
q^2(p-q))\nn
b&=&p q (\a\b p(\b-\d-p+2q)+\a\d q(\b-\d-2p+q)
-\b\d q(2p-q)\nn
&&-\a pq(p-q)-\d q^2(\d+p-q)) \nn
c&=& -\d q^2(p\b+q\d+pq)\ .
\label{gamma}
\eea
which allows us to readily express $\g$ as a function of the other
five parameters. Note however that the solutions of (\ref{gamma})
still have to be supplemented by the condition $\g >0$, which excludes
one of the two roots of \r{gamma}. The remaining one yields
\bea
\g&=&\frac{\g_1+\g_2+\g_3}{\g_4}\ ,\nn
\g_1&=&-p(-\a\b^2 p + \a\b\d p + \a\b p^2 - \a\b\d q + \a\d^2q -
2\a\b pq +2\a\d pq) \ ,\nn
\g_2&=&-p(2\b\d pq+\a p^2q-\a\d q^2 - \b\d q^2 +\d^2q^2 - \a pq^2 +\d
pq^2-\d q^3) \ ,\nn
\g_3&=& p(\a\b p + \a\d q + \a pq + \d q^2)
\sqrt{\b^2+2\b\d +\d^2-2\b p+2\d p+p^2+2q(\b-\d-p)+q^2}\nn
\g_4&=&2\d q(\b p + \d q + pq)\ .
\label{g}
\eea
For the special case $\g=\d=0$ (\ref{constr2}) has the simple solution
$\b=-q+\frac{pq}{\a+q}$. 

Using \r{C2} and \r{gamma} one can show that the region of the
phase diagram accessible by the 2-d representation is (without loss of
generality we assume $p\geq q$)
\be
\kappa_+(\b,\d)>\frac{1}{\kappa_+(\a,\g)}\ .
\label{region}
\ee
This can be easily checked for the case $\g=\d=0$, for the general
case we carried out a numerical analysis. The region described
by \r{region} covers the area above the dotted line in Fig. 2, {\sl
i.e.} most of the phases $A$ and $B$. We note that for the case of
symmetric diffusion $p=q$ the two-dimensional representation does not
exist. The infinite-dimensional representation is given by \r{71}.

\item[c)]{} Three-dimensional representation

An analogous analysis of the constraint $f_3=0$ for the
three-dimensional representation leads to the same constraint
\r{region}. We believe this to hold for any finite-dimensional
representation as well. 

\end{itemize}

\section{Matrix Elements of the Two-Dimensional Representation}
Using \r{bb},\r{a},\r{b} and \r{ini} and performing a similarity
transformation with 
\be
S=\left(\matrix{1&0\cr 0&\sqrt{\frac{x_6}{x_5}}\cr}\right) \ ,
\ee
we obtain the following form for the matrices $A$ and $B$ 
\be
A=\left(\matrix{0&f_1\cr 0&a_2\cr}\right)\ ,\ 
B=\left(\matrix{0&0\cr f_1&b_2\cr}\right)\ ,\ \v=\pmatrix{1\cr 0\cr}\
,\ \w =\pmatrix{1&0\cr}\ ,
\label{2d}
\ee
where
\be
f_1^2=\frac{x_7}{x_2}\ ,\ b_2=\frac{x_2x_5-x_1x_6}{x_2^2-x_1x_4}
\ ,\ a_2=\frac{x_2x_6-x_4x_5}{x_2^2-x_1x_4}\ .
\ee
Using the constraint $f_2=0$, $a_2$ and $b_2$ can be rewritten as {\sl
e.g.} 
\bea
a_2&=&\frac{(\a\b-\g\d)(p\b+q\d)-(p-q)[p\b(\a+\d)+q\d(\b+\g)]}
{p^2\a\b-q^2\g\d} \ ,\nn
b_2&=&\frac{(\a\b-\g\d)(p\a+q\g)-(p-q)[p\a(\b+\g)+q\g(\a+\d)]}
{p^2\a\b-q^2\g\d} \ .
\eea

For actual computations properties of the matrix
\be
C=D+E=\frac{\a+\b+\g+\d}{\a\b -\g\d} + \frac{\a+\g}{\a\b -\g\d} A
+ \frac{\b+\d}{\a\b -\g\d} B
\label{c}
\ee
are of central importance. We have to distinguish between two cases: 
if $\a\neq\b$ or $\g\neq\d$, $C$ can be diagonalized
\bea
SCS^{-1}&=& \left(\matrix{\la_+&0\cr 0&\la_-\cr}\right)\ ,\
S=\frac{1}{\a\b-\g\d}\left(\matrix{(\b+\d)f_1&(\a+\g)a_2\cr
(\b+\d)f_1 &(\b+\d)b_2\cr}\right)\ , \nn
\la_+&=&\frac{\a+\b+\g+\d}{\a\b -\g\d} + \frac{\a+\g}{\a\b -\g\d} a_2\
,\nn \la_-&=&\frac{\a+\b+\g+\d}{\a\b -\g\d} + \frac{\b+\d}{\a\b -\g\d}
b_2\ .
\label{la+-}
\eea
Note that the determinant $\det S=(\b+\d)f_1(\la_--\la_+)$ is
different from zero unless $\la_-\equiv\la_+$ (the prefactor vanishes
only if $x_7=0$, in which case the 2-d representation breaks up into
1-d representations).
In the case $\la_+=\la_-$, $C$ can no longer be diagonalized but only
be brought to Jordan normal form. The condition $\la_+=\la_-$ can be
rewritten as
\be
0=\frac{(\a+\g)a_2-(\b+\d)b_2}{\a\b-\g\d}=
\frac{p-q}{p^2\a\b-q^2\g\d}[\b\g-\a\d-p(\a-\b)-q(\d-\g)]\ . 
\ee
As the 2-d representation does not exist for $p=q$ we conclude that
the only case, in which $\la_+=\la_-$ is if
\be
\g=\frac{\a\d+p(\a-\b)+q\d}{\b+q}\ .
\label{g2}
\ee
A numerical analysis of the two conditions \r{g} and \r{g2} yields
that a necessary condition for $\la_+=\la_-$ is that $\a=\b$ and
$\g=\d$, {\sl i.e.} the coexistence line of Fig.2.

\section{Correlation Functions Off the Coexistence Line
($\la_+\neq\la_-$)} 

The density profile $\langle\tau_j\rangle$ is readily evaluated using
\r{jttt},\r{c} and \r{la+-}
\bea
\langle \tau_j\rangle&=& \frac{\w C^{j-1}DC^{L-j}\v}{\w C^{L}\v}
=\frac{\w S^{-1}SC^{j-1}S^{-1}SDS^{-1}SC^{L-j}S^{-1}S\v}
{\w S^{-1}SC^{L}S^{-1}S\v}\nn
&=&\frac{\wt \left(\matrix{\la_+^{j-1}&0\cr 0&\la_-^{j-1}\cr}\right)
{\tilde D}\left(\matrix{\la_+^{L-j}&0\cr 0&\la_-^{L-j}\cr}\right)\vt}
{\wt \left(\matrix{\la_+^{L}&0\cr 0&\la_-^{L}\cr}\right)\vt}\ ,
\eea
where $\vt=\frac{(\b+\d)f_1}{\a\b-\g\d}\pmatrix{1\cr 1}$,
$\wt=\frac{(\a\b-\g\d)}{f_1(\b+\d)[(\a+\g)a_2-(\b+\d)b_2]}
\pmatrix{-(\b+\d)b_2\ ,&(\a+\g)a_2}$, and  
\be
{\tilde D}=SDS^{-1}=\left(\matrix{\frac{\a+\d+\a a_2}{\a\b-\g\d}
&\frac{-a_2}{(\b+\d)}\cr 0&
\frac{\a+\d+\d b_2}{\a\b-\g\d}\cr}\right).
\ee
Note that one of the matrix elements of $\tilde D$ vanishes. This will
have consequences for the shape of the density profile and the
two-point function.

The scalar products are easy to work out and lead to the following
result for the density profile
\be
\langle\tau_j\rangle =\Omega\left(\omega_0 +\omega_1
\exp\left(\frac{L-j}{\zeta}\right)+\omega_2
\exp\left(\frac{L-1}{\zeta}\right) \right)\ ,
\ee
where
\bea
\Omega &=&\frac{1}{[\a+\b+\g+\d+(\a+\g)a_2][(\b+\d)b_2-(\a+\g)a_2
\exp\left(\frac{L}{\zeta}\right)]}\nn
\frac{1}{\zeta}&=&\ln\left(\frac{\la_-}{\la_+}\right)=
\ln\left(\frac{\a+\b+\g+\d+(\b+\d)b_2}
{\a+\b+\g+\d+(\a+\g)a_2}\right)\ ,\nn
\omega_0&=& (\b+\d)b_2(\a+\d+\a a_2)\ ,\nn
\omega_1&=& -a_2b_2\ (\a\b-\g\d)\ ,\nn
\omega_2&=& -(\a+\g)a_2(\a+\d+\d b_2)\ .
\eea
Two cases have to be distinguished: 
\begin{itemize}
\item{} $\zeta< 0$, which corresponds to the case $\kappa_+(\a,\g)
>\kappa_+(\b,\d)$ {\sl i.e.} Phase B. In this case the profile for
$L\gg\zeta$ is of the form
\be
\langle\tau_j\rangle = m_<+c_< \exp\left(\frac{j-L}{|\zeta|}\right)
+{\cal O}\left(\exp\left(-\frac{L}{|\zeta|}\right)\right)\ ,
\ee
where
\bea
m_<&=&\frac{\a+\d+\a a_2}{\a+\b+\g+\d+(\a+\g)a_2}\ ,\nn
c_<&=&\frac{-a_2(\a\b-\g\d)}{[\a+\b+\g+\d+(\a+\g)a_2](\b+\d)}\ >0.
\label{m1}
\eea
The average density starts at the value $m_<$ at the left boundary,
remains constant throughout the bulk, and eventually exhibits an
exponential increase to the value $m_<+c_<$ at the right
boundary.

\item{} $\zeta >0$, which corresponds to the case $\kappa_+(\a,\g)
<\kappa_+(\b,\d)$ {\sl i.e.} Phase A. In this case the profile for
$L\gg\zeta$ is of the form
\be
\langle\tau_j\rangle = m_>+c_> \exp\left(-\frac{j-1}{|\zeta|}\right)
+{\cal O}\left(\exp\left(-\frac{L-1}{|\zeta|}\right)\right)\ ,
\ee
where
\bea
m_>&=&\frac{\a+\d+\d b_2}{\a+\b+\g+\d+(\b+\d)b_2}\ ,\nn
c_>&=&\frac{b_2(\a\b-\g\d)}{[\a+\b+\g+\d+(\b+\d)b_2](\a+\g)}<0\ .
\label{m2}
\eea
Here the density starts at the value $m_>+c_>\exp(-\frac{1}{\zeta})$
at the left boundary, increases exponentially to $m_>$, and remains
constant until the right boundary.
\end{itemize}

Let us now take a closer look at the expressions for the bulk
densities $m_<$ and $m_>$ and the correlation length $|\zeta|$. It
turns out that they are ``{\sl universal}'' in the sense that they
depend only on the two variables $\kappa_+(\a,\g)$ and
$\kappa_+(\b,\d)$ instead of all five independent parameters
$\a,\b,\d,p,q$. We start with $m_<$ and $m_>$. One can show that  
\bea
m_<&=&\frac{1}{1+\kappa_+(\a,\g)}\ ,\nn
m_>&=&\frac{\kappa_+(\b,\d)}{1+\kappa_+(\b,\d)}\ .
\label{magn}
\eea
The equality of \r{magn} with \r{m1} and \r{m2} is established
analytically only for $0=\g=\d$, and numerically to machine accuracy
for the general case. As we will now argue, we believe \r{magn} to hold
not only for the 2-d representation but in general for phases $A$ and
$B$: in Appendix F a mean-field analysis of the partially asymmetric
diffusion process is carried out, and $m_<$ and $m_>$ in phases B and
A are determined. We denote the corresponding results (see
Appendix D) by $m_{<,MF}$ and $m_{>,MF}$. It is straightforward to
demonstrate analytically that 
\be
m_{<,MF}=\frac{1}{1+\kappa_+(\a,\g)}\ ,\
m_{>,MF}=\frac{\kappa_+(\b,\d)}{1+\kappa_+(\b,\d)}\ .
\ee
This shows that the mean-field theory result is universal in phases A and
B, and in addition is exact whenever the 2-d representation exists.
Based on this observation and the results of \cite{dehp} for
$0=\g=\d=q$ (see eqn \r{asdens}) we conjecture that \r{magn} holds
true throughout phases A and B.

The value of $m_>$ amd $m_<$ as a function of $\kappa_+(\a,\g)$ and
$\kappa_+(\b,\d)$ is shown in Fig.3. Accordingly phase A is identified
as a {\sl high-density phase} with $m_>> \frac{1}{2}$, and phase B as a
{\sl low-density phase} with $m_<< \frac{1}{2}$.
\begin{figure}[htbtv]
\begin{center}
\setlength{\epsfysize}{100mm}
\setlength{\epsfxsize}{100mm}
\leavevmode
\epsfbox{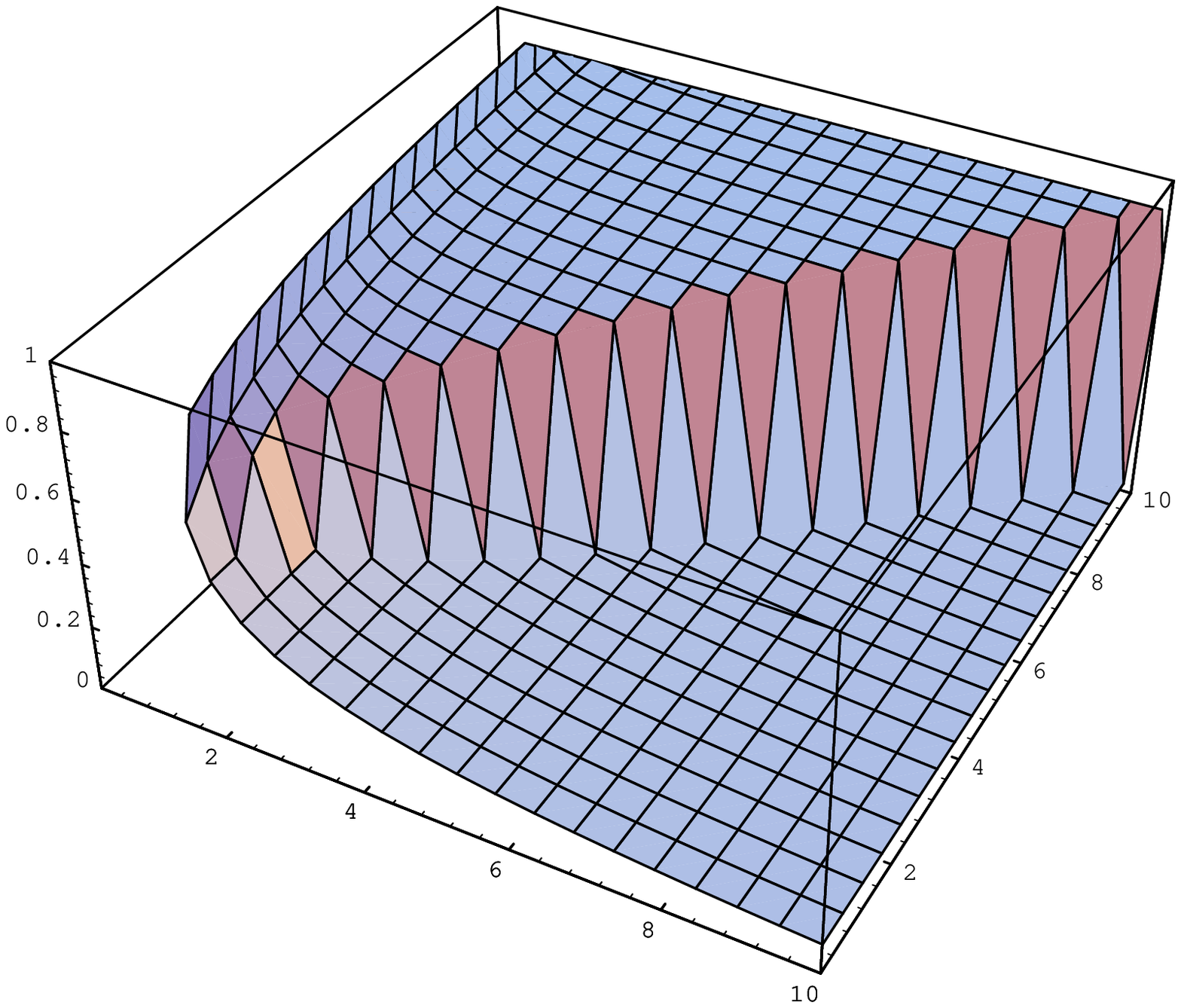}
\end{center}
\caption{\label{dens}}
{\sl Plot of the bulk densities $m_>$, $m_<$ in phases A and B in the
region accessible by the 2-d representation.}
\vskip -4.5cm
\hskip 4.8cm$\kappa_+(\a,\g)$
\vskip -1.5cm
\hskip 12cm $\kappa_+(\b,\d)$
\vskip -3cm \hskip 1.1cm $m_>,m_<$
\vskip 7cm
\end{figure}

Unlike the quantities $m_>$ and $m_<$ the coefficients $c_<$ and $c_>$
of the exponentials are {\sl not} universal in the sense that they are
not only functions of $\kappa_+(\a,\g)$ and $\kappa_+(\b,\d)$.
However, the correlation length $|\zeta|$ in the exponential can be
expressed as
\be
\exp\left(\frac{1}{\zeta}\right)=\frac{\kappa_+(\a,\g)}{\kappa_+(\b,\d)}
\left(\frac{1+\kappa_+(\b,\d)}{1+\kappa_+(\a,\g)}\right)^2\ .
\label{corrl}
\ee
This shows that the correlation length $|\zeta|$ diverges when
$\kappa_+(\a,\g)$ and $\kappa_+(\b,\d)$ approach the coexistence line:
$\zeta\rightarrow\infty$ when approaching the coexistence line from
phase $A$, and $\zeta\rightarrow -\infty$ from phase $B$.
It is interesting to compare \r{asdens} with \r{corrl}. Surprisingly
enough, for phases $A_I$ and $B_I$ the $0=\g=\d=q$ correlation length
has precisely the expression \r{corrl}. This ceases to be the case for
phases $A_{II}$ and $B_{II}$. We would like to stress that although
the mean-field values \r{magn} are exact, the correlation length
$\zeta$ given by \r{corrl} cannot be obtained in the mean-field
approximation. 
\vskip .5cm 

The two-point function can be evaluated in way analogous to the case
of the one-point function discussed above. After some straightforward
computations we obtain
\be
\langle\tau_j\tau_k\rangle =\Omega'\left(\omega_3 +\omega_4 
\exp\left(\frac{L-k}{\zeta}\right)+\omega_5
\exp\left(\frac{L-j-1}{\zeta}\right) +\omega_6
\exp\left(\frac{L-2}{\zeta}\right)\right)\ , 
\ee
where
\bea
\Omega'&=&\frac{1}{\la_+^2[(\b+\d)b_2-(\a+\g)a_2
\exp\left(\frac{L}{\zeta}\right)]}\ ,\nn
\omega_3&=& (\b+\d)b_2\left(\frac{\a+\d+\a a_2}{\a\b-\g\d}\right)^2\ ,\nn
\omega_4&=& -a_2b_2\ \frac{\a+\d+\a a_2}{\a\b-\g\d}\ ,\nn
\omega_5&=& -a_2b_2\ \frac{\a+\d+\d b_2}{\a\b-\g\d}\ ,\nn
\omega_6&=& -(\a +\g)a_2\left(\frac{\a+\d+\d b_2}{\a\b-\g\d}\right)^2\ .
\eea
Again we have to distinguish between two cases 
\begin{itemize}
\item{} $\zeta <0$ (low-density phase):\ 
The connected two-point function in the large $L$ limit is given by
\be
\langle\tau_j\tau_k\rangle-\langle\tau_j\rangle\langle\tau_k\rangle
=c_{<}(m_>-m_<)\exp\left(\frac{L-j}{\zeta}\right)
-c_{<}^2 \exp\left(\frac{2L-j-k}{\zeta}\right)\ ,
\label{2pB}
\ee
It is different from zero only very close to the right boundary, from
where it decays exponentially. 

\item{} $\zeta >0$ (high-density phase) :
\be
\langle\tau_j\tau_k\rangle-\langle\tau_j\rangle\langle\tau_k\rangle
=c_{>}(m_<-m_>)\exp\left(\frac{-k+1}{\zeta}\right)
-c_{>}^2 \exp\left(\frac{-j-k+2}{\zeta}\right)\ ,
\label{2pA}
\ee
Thus the connected two-point function in phase A is different from zero
only very close to the left boundary, where it exhibits an exponential
behaviour.

\end{itemize}

\section{Correlation Functions on the Coexistence Line ($\la_+=\la_-$)}

For the case $\a=\b$, $\g=\d$ we have $\kappa_+(\a ,\g)=\kappa_+(\b
,\d)>1$ with
\be
\kappa_+(\a,\g)=\sqrt{\frac{p}{q}}
\ee
and are thus on the phase boundary between the high-density
phase $A$ and the low-density phase $B$. Here we have taken into
account the constraint \r{gamma} for the 2-d representation, which can
be solved with the result 
\be
\g=-p+(\a+q)\sqrt{\frac{p}{q}}\ .
\ee
The matrix elements of the matrices $A$ and $B$ in \r{2d}  are found to
be 
\be
a_2=b_2=-if_1=\sqrt{\frac{p}{q}}-1-\a\frac{\sqrt{q}+\sqrt{p}}
{q\sqrt{p}}\ ,
\ee 
which leads to the following form for the matrix $C$
\be
C= \frac{1}{\a-\g}\left((2+a_2)I+a_2 P\right)\ ,\ 
P=\left(\matrix{-1& i\cr i&1\cr}\right) \ .
\ee
The matrix $P$ has the property $P^2=0$, which is important for carrying
out the calculations below.
Using the fact that $P^2=0$ it is easy to show that
\be
C^k=\frac{(2+a_2)^{k-1}}{(\a-\g)^k}\left(\matrix{
2+(1-k)a_2&ika_2\cr ika_2&2+(1+k)a_2\cr}\right)\ ,
\label{ck}
\ee
which implies that the normalization is given by
\be
\w C^L\v =\frac{(2+a_2)^{L-1}}{(\a-\g)^L} (2+(1-L)a_2)\ .
\label{norm}
\ee 
Using (\ref{norm}) we can easily evaluate the current
\be
J=\left(\frac{\a-\g}{2+a_2}\right)
\left(\frac{2+(2-L)a_2}{2+(1-L)a_2}\right) .
\ee
In the thermodynamic limit this simplifies to
\be
J=\frac{\a-\g}{2+a_2}=\sqrt{pq}\frac{1-{\cal Q}}
{1+{\cal Q}}\ ,\quad {\cal Q}={\sqrt{\frac{q}{p}}}\equiv
\frac{1}{\kappa_+(\a,\g)}\ , 
\ee
which in turn can be shown to be equal to \r{C10}.

The density profile is
\be
\langle\tau_j\rangle=
\frac{(\a+\g)(2+a_2(2-L))+a_2^2(\g-L\a)}{(\a+\g)(2+a_2)(2+(1-L)a_2)}
+\frac{(\a-\g)a_2^2}{(\a+\g)(2+a_2)(2+(1-L)a_2)} j\ .
\ee
In the thermodynamic limit $j,L\rightarrow\infty$ $\frac{j}{L}=x$
fixed, this turns into
\bea
\langle\tau_{Lx}\rangle&=&
\frac{{\cal Q}}{1+{\cal Q}}+
\frac{{1-\cal Q}}{1+{\cal Q}} x\ .
\label{tjII}
\eea
This means that for $p>q$ the density increases linearly from
$\frac{{\cal Q}}{1+{\cal Q}}$ at the left end of the chain to
$\frac{1}{1+{\cal Q}}$, whereas for $q>p$ is decreases linearly from
$\frac{{\cal Q}}{1+{\cal Q}}$ to $\frac{1}{1+{\cal Q}}$. Clearly the
profile is symmetric under simulataneous interchange of $p$ and $q$
and left and right as it should be. The most remarkable feature of the
profile \r{tjII} is the fact that it is independent of the
injection/extraction rate $\a$. The only relevant parameter is the
ratio $\frac{p}{q}$ of the diffusion rates to the right and to the
left. This fact is probably a feature of the particular
representation we work with since in the completely asymmetric case the
current and density profile on the coexistence line {\sl are}
dependent on the boundary condition $\a$ (see \r{leading}), and we
expect that in general the current and density profile in the
partially asymmetric case will depend on the boundary conditions as
well. 

The two-point function can be determined by using \r{ck} in \r{jttt}
and is found to be of the form
\be
\langle\tau_j\tau_k\rangle =\omega_7 +\omega_8 j+\omega_9 k\ ,
\ee
where
\bea
\omega_7&=&
\frac{(\a+\g)^2[2+a_2(3-L)+a_2^2]+2a_2^2[\g^2-L\a(\a+\g)]+a_2^3[\g^2-L\a^2]}
{(2+a_2)^2(2+a_2(1-L))(\a+\g)^2}\ , \nn
\omega_8&=&\frac{(\a-\g)a_2^2(\a+\g+\g a_2)}
{(2+a_2)^2(2+a_2(1-L))(\a+\g)^2}\ , \nn
\omega_9&=&\frac{(\a-\g)a_2^2(\a+\g+\a a_2)}
{(2+a_2)^2(2+a_2(1-L))(\a+\g)^2}\ .
\eea

In the limit $j,k,L\rightarrow\infty$ with $\frac{j}{L}=x$ and
$\frac{k}{L}=y$ fixed this simplifies essentially
\be
\langle\tau_{Lx}\tau_{Ly}\rangle =\left(\frac{\cq}{1+\cq}\right)^2
+\frac{1-\cq}{\left(1+\cq\right)^2} x
+\frac{\cq(1-\cq)}{\left(1+\cq\right)^2} y \ .
\ee
Using \r{tjII} we finally arrive at the following result for the
connected two-point function
\be
\langle\tau_{Lx}\tau_{Ly}\rangle -\langle\tau_{Lx}\rangle
\langle\tau_{Ly}\rangle = \left(\frac{1-\cq}{1+\cq}\right)^2 x(1-y)\ .
\label{2pcoex}
\ee
Like the density profile the connected two-point function is
independent of $\alpha$. It takes its maximal value
$\left(\frac{1-\cq}{2(1+\cq)}\right)^2$ in the middle of the chain, and
decreases to zero at both boundaries. It is interesting to note that a
similar expression for the connected two-point function was obtained
for the completely asymmetric (particles hop only to the right) {\sl
deterministic} exclusion model with stochastic boundary effects
\cite{schuetz}. 

\section{Summary and Conclusion}
In this paper we have first presented the Fock representations of the
quadratic algebra. The representations can be either
infinite-dimensional or finite-dimensional. Each finite-dimensional
representation is characterized by a constraint on the seven
parameters of the algebra. The matrix elements of the two generators
of the algebra are given by recurrence relations. Only in
the cases where these relations can be solved in a simple way the
corresponding Fock-representation is useful for studying the physical
problem of partially asymmetric diffusion with open boundaries.
Quadratic algebras with more than two generators, which are relevant
for many-state problems, will be presented elsewhere. The structure of
the associative algebra is different for these cases \cite{multi}. The
boundary conditions define representations of a different type as
compared to the Fock representations considered here.

The quadratic algebra appears in the DEHP Ansatz for the steady-state
probability distributions of one-dimensional reaction diffusion
problems with two-body rates and injection and extraction of particles
at the ends of the chain. As shown in section V and also in
\cite{shocks,godreche}, the quadratic algebra appears also for some more
general reaction-diffusion processes. As also shown in the present
work, if one considers three-body rates, cubic algebras occur. We have
not studied the representation theory for that case. 

We have used the 2-d representation of the quadratic algebra to
compute the density profile and correlation functions. Both have a
special dependence on the coordinates (see \r{2pA} and \r{2pB}). The
parameter dependence is also peculiar. Certain quantities like the
density around the middle of the chain or the correlation length
depend on the parameters $\a,\b,\g,\d,p,q$ only through two functions
$\kappa_+(\a,\g)$ and $\kappa_+(\b,\d)$ defined in the text. Our
results and those of \cite{ddm,dehp,sd} suggest that the density
in the bulk is given by the mean-field prediction
in the thermodynamic limit. Again from our results and those of
\cite{ddm,dehp,sd} it looks like the correlation lengths again are
``universal'' (they depend only on $\kappa_+(\a,\g)$ and
$\kappa_+(\b,\d)$) in regions $A_I$ and $B_I$ of the phase diagram.

By means of \r{trans} our results \r{2pB}, \r{2pA} and \r{2pcoex} for
the connected two-point functions can be directly translated to the
XXZ quantum-spin chain \r{A5d} with non-diagonal boundary terms.
This yields new and nontrivial results for correlators in the XXZ
chain with boundaries. 
We have not considered the problem of time-dependent correlation
functions. Some recent numerical results of Bilstein \cite{u} can give
a hint in this direction. He studied the finite-size scaling bahaviour
of the energy gaps of the quantum hamiltonian. It turns out that in
the low- and high-density phases the system is massive. In phase $C$
(see Fig.2) the system is massless. Both real and imaginary parts of
the energy gaps vanish like $L^{-\frac{3}{2}}$, where $L$ is the size
of the system. The coexistence line is also massless but features a
less simple length dependence.

Finally, as a by-product of our work on the DEHP construction, we have
given a simple way to construct irreducible representations of the
$U_q(SU(2))$ quantum group.

\vspace*{1cm}

\begin{center}
{\large \sc Acknowledgements}
\end{center}

We are grateful to A. Berkovich for a useful suggestion and to K.
Krebs, M. Scheunert, H.Z. Simon and G. Sch\"utz for reading the
manuscript and discussions. We thank A. Dzhumadildaev for pointing out
refererences \cite{words} and \cite{multi} to us.
\vspace*{2cm}

\appendix{\Large\bf Appendix A}
\vspace*{.5cm}

We would like to show that in the Fock representation of the quadratic
algebra \r{ab} the matrices $A$ and $B$ can be written in tridiagonal
form. This is a property of the Fock representation only, and is
independent of the parameters $x_i$. Instead of presenting the general
proof, we give two examples which will help the reader to see the
mechanism behind the proof. We start with the 2-d representation. 
We choose $\v=\pmatrix{1\cr 0}$ and take
$\w=\pmatrix{w_1, & w_2}$ with $w_1\neq 0$ such that $\langle
W|V\rangle\neq 0$. We then perform a similarity transformation
\be
S=\left(\matrix{1&\a\cr 0&1}\right)\ ,\ S\v=\v\ ,\ \w
S^{-1}=\pmatrix{w_1,& -\a w_1+w_2}\ ,
\ee
where we choose $\a=\frac{w_2}{w_1}$ and $w_1=1$, which results in
$\w=\pmatrix{1 ,& 0}$ and $\langle W|V\rangle= 1$.
In this basis $A$ and $B$ now must be of the form
\be
A=\left(\matrix{0&a_{12}\cr 0&a_{22}\cr}\right)\ ,\
B=\left(\matrix{0&0\cr b_{21}&b_{22}\cr}\right)\ .
\ee
We now perform another similarity transformation
\be
S=\left(\matrix{1&0\cr 0&\b\cr}\right)\ ,\ \w S^{-1}=\w\ ,\ S\v=\v
\ee
where we choose $\b^2=\frac{a_{12}}{b_{21}}$ and find two equivalent
representations 
\be
A=\left(\matrix{0&\pm\sqrt{a_{12}b_{21}}\cr 0&a_{22}\cr}\right)\ ,\
B=\left(\matrix{0&0\cr \pm\sqrt{a_{12}b_{21}}&b_{22}\cr}\right)\ .
\label{absim}
\ee
Here we have assumed that both $a_{12}$ and $b_{21}$ are different
from zero. One can show that if $a_{12}=0$ or $b_{21}=0$ the algebra
\r{ab} yields the condition $x_7=0$, which corresponds to the
1-d representation and thus the 2-d representation is not
interesting. Renaming the entries of $A$ and $B$ we arrive at
\be
A=\left(\matrix{0&f_1\cr 0&a_{2}\cr}\right)\ ,\
B=\left(\matrix{0&0\cr f_1&b_{2}\cr}\right)\ .
\ee

Let us now consider the 3-d representation. We can repeat the
similarity transformations from the 2-d case to bring $A$ and $B$ to
the form ($f_1'\neq 0$) 
\be
A=\left(\matrix{0&f_1'&g_1'\cr 0&a_{22}'&a_{23}'\cr
0&a_{32}'&a_{33}'\cr}\right)\ ,\ 
B=\left(\matrix{0&0&0\cr f_1'&b_{22}'&b_{23}'\cr
g_1'&b_{32}'&b_{33}'\cr}\right)\ , 
\ee
where $\v=\pmatrix{1\cr 0\cr 0\cr}$ and $\w=\pmatrix{1,& 0,& 0}$.
Taking the further similarity transformation
\be
S=\left(\matrix{1&0&0\cr0&\cos(\theta)&\sin(\theta)\cr
0&-\sin(\theta)&\cos(\theta)\cr}\right)\ ,\quad
\tan(\theta)=\frac{g_1'}{f_1'}
\ee
we get
\be
A=\left(\matrix{0&f_1&0\cr 0&a_{22}&a_{23}\cr
0&a_{32}&a_{33}\cr}\right)\ ,\ 
B=\left(\matrix{0&0&0\cr f_1&b_{22}&b_{23}\cr
0&b_{32}&b_{33}\cr}\right)\ .
\ee
Finally, by means of a diagonal similarity transformation we can bring
$A$ and $B$ to the form
\be
A=\left(\matrix{0&f_1&0\cr 0&a_{2}&f_{2}\cr
0&h_{2}&a_{3}\cr}\right)\ ,\ 
B=\left(\matrix{0&0&0\cr f_1&b_{2}&k_{2}\cr
0&f_{2}&b_{3}\cr}\right)\ .
\ee

This procedure can be generalized to any dimension of the
representation. We can thus search for representations of the
quadratic algebra starting with the tridiagonal form

\be
\tA=\left(\matrix{0&f_1&0&0&0&0&0\ldots\cr
0&a_2&f_2&0&0&0&0\ldots\cr
0&h_2&a_3&f_3&0&0&0\ldots\cr
0&0&h_3&a_4&f_4&0&0\ldots\cr
\ldots& & & & \ldots& &\cr}\right)
\ee

\be
\tB=\left(\matrix{0&0&0&0&0&0&0\ldots\cr
f_1&b_2&k_2&0&0&0&0\ldots\cr
0&f_2&b_3&k_3&0&0&0\ldots\cr
0&0&f_3&b_4&k_4&0&0\ldots\cr
\ldots& & & &\ldots& &\cr}\right)\ .
\ee

\vspace*{.5cm}
\appendix{\Large\bf Appendix B}
\vspace*{.5cm}

In this appendix we discuss a generalization of the DEHP-formalism to
chemical processes of the types given in \r{rates} that incorporate
{\sl three} neighbouring sites. By construction all models discussed
above are contained in the present formulation.
From a physical point of view this generalization is quite interesting. 
For purely diffusive processes it allows us {\sl e.g.} to let particles 
hop to the next-nearest neighbour site if the nearest neighbour site is
unoccupied. In the corresponding traffic-flow picture this corresponds
to letting cars move faster or slower depending on whether the road ahead 
is free for a long or a short distance.

We consider a master-equation of the form
\bea
0=\frac{\partial P}{\partial
t}&=&-\sum_{k=1}^{L-2}\sum_{\g_k,\g_{k+1},\g_{k+2}}
\left(H_{k,k+1,k+2}\right)_{\tau_k,\tau_{k+1},\tau_{k+2}
}^{\g_k,\g_{k+1},\g_{k+2}}P_L(\tau_1,\tau_2\ldots\tau_{k-1},
\g_k,\g_{k+1},\g_{k+2},\tau_{k+3}\ldots\tau_L)\nn 
&&-\sum_{\g_1\g_2}(h_{12})_{\tau_1\tau_2}^{\g_1\g_2}
P_L(\g_1\g_2\tau_3\ldots\tau_L)
-\sum_{\g_{L-1}\g_L}(h_{L-1L})_{\tau_{L-1}\tau_{L}}^{\g_{L-1}\g_L}
P_L(\tau_1\ldots\tau_{L-2}\g_{L-1}\g_{L} )\ ,
\label{3meq}
\eea
where
\be
\left(H_{k,k+1,k+2}\right)_{\tau_k,\tau_{k+1},
\tau_{k+2}}^{\g_k,\g_{k+1},\g_{k+2}}
=\Bigg\lbrace\matrix{\sum'_{\b_k,\b_{k+1},\b_{k+2}}
\Gamma_{\b_k,\b_{k+1},\b_{k+2}}^{\g_k,\g_{k+1},\g_{k+2}}\qquad
\g_j=\tau_j\ ,\ j=k,k+1,k+2\cr
-\Gamma_{\tau_k,\tau_{k+1},\tau_{k+2}}^{\g_k,\g_{k+1},\g_{k+2}}\qquad
{\rm else\ . }\cr}
\ee
Here $\G^{\a\b\g}_{\a'\b'\g'}$ is the probability per unit time that
the configuration $(\a\b\g)$ on three neighbouring sites changes to the 
configuration $(\a'\b'\g')$.
Following the analysis in section $1$ above we demand for the probability 
distribution in the bulk that
\bea
-\sum_{\g_k,\g_{k+1},\g_{k+2}}&&\left(H_{k,k+1,k+2}\right)_{
\tau_k,\tau_{k+1},\tau_{k+2}}^{\g_k,\g_{k+1},\g_{k+2}}
P_L(\tau_1,\tau_2\ldots\tau_{k-1},\g_k,\g_{k+1},\g_{k+2},\tau_{k+3}
\ldots\tau_L)\nn 
=&& x_{\tau_k} P_{L-1}(\tau_1\ldots\tau_{k-1}\tau_{k+1}\ldots \tau_L)
 -x_{\tau_{k+1}} P_{L-1}(\tau_1\ldots\tau_{k}\tau_{k+2}\ldots \tau_L)\nn
&&+ y_{\tau_k} P_{L-1}(\tau_1\ldots\tau_{k-1}\tau_{k+1}\ldots \tau_L)
 -y_{\tau_{k+2}} P_{L-1}(\tau_1\ldots\tau_{k+1}\tau_{k+3}\ldots \tau_L)\nn
&&+ z_{\tau_{k+1}} P_{L-1}(\tau_1\ldots\tau_{k}\tau_{k+2}\ldots \tau_L)
 -z_{\tau_{k+2}} P_{L-1}(\tau_1\ldots\tau_{k+1}\tau_{k+3}\ldots \tau_L)\nn
&&+ t_{\tau_k\tau_{k+1}}
P_{L-2}(\tau_1\ldots\tau_{k-1}\tau_{k+2}\ldots \tau_L) 
-t_{\tau_{k+1}\tau_{k+2}}
P_{L-2}(\tau_1\ldots\tau_{k}\tau_{k+3}\ldots \tau_L) . 
\label{3x}
\eea
This ensures that there will be no contribution from the bulk to the
r.h.s. of \r{3meq}, as all terms will cancel when summed over 
$(k,k+1,k+2)$. As compared to \r{stat1} it is now possible to include
terms containing probability distributions $P_{L-2}$ with two fewer
particles. The r.h.s. of \r{3x} can be simplified by introducing the
notation
\be
\mu_{b_k}=x_{b_k}+y_{b_k}\ ,\ \nu_{b_k}=z_{b_k}-x_{b_k}\ ,\
\lambda_{b_k}=-\mu_{b_k}-\nu_{b_k}\ .
\ee
From the boundaries we get the conditions that
\bea
&&\sum_{\g_1\g_2}(h_{12})_{\tau_1\tau_2}^{\g_1\g_2}
P_L(\g_1\g_2\tau_3\ldots\tau_L)=t_{\tau_1\tau_2}
P_{L-2}(\tau_3\ldots\tau_L)+\mu_{\tau_1}P_{L-1}(\tau_2\ldots\tau_L)\nn
&&\hskip 4cm-\la_{\tau_2}P_{L-1}(\tau_1\tau_3\ldots\tau_L)\nn
&&\sum_{\g_{L-1}\g_L}(h_{L-1L})_{\tau_{L-1}\tau_{L}}^{\g_{L-1}\g_L}
P_L(\tau_1\ldots\tau_{L-2}\g_{L-1}\g_{L} )=\nn
&&-t_{\tau_{L-1}\tau_L}P_{L-2}(\tau_1\ldots\tau_{L-2})
-\mu_{\tau_{L-1}}P_{L-1}(\tau_1\ldots\tau_{L-2}\tau_L)
+\la_{\tau_L}P_{L-1}(\tau_1\ldots\tau_{L-1})\ .
\eea
Here we have allowed for arbitrary processes of the type introduced in
\r{rates} to occur on the first two and last two sites of the lattice.
Putting everything together we arrive at the following algebra
\be
-{\cal H}\pmatrix{D^3\cr D^2E\cr DED\cr ED^2\cr DE^2\cr EDE\cr E^2D\cr
E^3\cr}= 
\pmatrix{
0\cr 
\mu_1DE+\nu_1DE+\la_0D^2+t_{11}E-t_{10}D\cr
\mu_1ED+\nu_0D^2+\la_1DE+t_{10}D-t_{01}D\cr
\mu_0D^2+\nu_1ED+\la_1ED+t_{01}D-t_{11}E\cr
\mu_1E^2+\nu_0DE+\la_0DE+t_{10}E-t_{00}D\cr
\mu_0DE+\nu_1E^2+\la_0ED+t_{01}E-t_{10}E\cr
\mu_0ED+\nu_0ED+\la_1E^2+t_{00}D-t_{01}E\cr
0\cr}\ ,
\ee
where
\be
{\cal H}=
\left(
\matrix{
H^{111}_{111}&\G^{110}_{111}&\G^{101}_{111}&\G^{011}_{111}&
\G^{100}_{111}&\G^{010}_{111}&\G^{001}_{111}&\G^{000}_{111}\cr
\G^{111}_{110}&H^{110}_{110}&\G^{101}_{110}&\G^{011}_{110}&
\G^{100}_{110}&\G^{010}_{110}&\G^{001}_{110}&\G^{000}_{110}\cr
\G^{111}_{101}&\G^{110}_{101}&H^{101}_{101}&\G^{011}_{101}&
\G^{100}_{101}&\G^{010}_{101}&\G^{001}_{101}&\G^{000}_{101}\cr
\G^{111}_{011}&\G^{110}_{011}&\G^{101}_{011}&H^{011}_{011}&
\G^{100}_{011}&\G^{010}_{011}&\G^{001}_{011}&\G^{000}_{011}\cr
\G^{111}_{100}&\G^{110}_{100}&\G^{101}_{100}&\G^{011}_{100}&
H^{100}_{100}&\G^{010}_{100}&\G^{001}_{100}&\G^{000}_{100}\cr
\G^{111}_{010}&\G^{110}_{010}&\G^{101}_{010}&\G^{011}_{010}&
\G^{100}_{010}&H^{010}_{010}&\G^{001}_{010}&\G^{000}_{010}\cr
\G^{111}_{001}&\G^{110}_{001}&\G^{101}_{001}&\G^{011}_{001}&
\G^{100}_{001}&\G^{010}_{001}&H^{001}_{001}&\G^{000}_{001}\cr
\G^{111}_{000}&\G^{110}_{000}&\G^{101}_{000}&\G^{011}_{000}&
\G^{100}_{000}&\G^{010}_{000}&\G^{001}_{000}&H^{000}_{000}\cr
}\right)\ .
\label{3genalg}
\ee
It is clear that only seven of the eight equations in (\ref{3genalg})
are independent. The boundary conditions are rewritten as
\bea
&&\sum_{\g_1\g_2}(h_{12})_{\tau_1\tau_2}^{\g_1\g_2}
\w[\g_1D+(1-\g_1)E][\g_2D+(1-\g_2)E]\nn
&&\qquad=
\w\bigg[(t_{\tau_1\tau_2}+\mu_{\tau_1}[\tau_2D+(1-\tau_2)E]-\la_{\tau_2}
[\tau_1D+(1-\tau_1)E]\bigg]\nn
&&\sum_{\g_{L-1}\g_L}(h_{L-1L})_{\tau_{L-1}\tau_L}^{\g_{L-1}\g_L}
[\g_{L-1}D+(1-\g_{L-1})E][\g_LD+(1-\g_L)E]\v\nn
&&\qquad=\bigg[-t_{\tau_{L-1}\tau_L}+\la_{\tau_L}[\tau_{L-1}D+
(1-\tau_{L-1})E]-\mu_{\tau_{L-1}} [\tau_LD+(1-\tau_L)E]\bigg]\v\ .
\eea
Note that the algebra \r{3genalg} is cubic and that the conditions at
the boundaries are quadratic in $D$ and $E$.

If we constrain ourselves to diffusion processes only, the system
(\ref{3genalg}) decouples into two sets of respectively two
independent equations  
\bea
{\cal H}_1\pmatrix{D^2E\cr DED\cr ED^2\cr} &=& 
\pmatrix{-\la_1DE+\la_0D^2+t_{11}E-t_{10}D\cr
\mu_1ED+\nu_0D^2+\la_1DE+t_{10}D-t_{01}D\cr}\ ,\nn
{\cal H}_2\pmatrix{DE^2\cr EDE\cr E^2D\cr} &=& 
\pmatrix{\mu_1E^2-\mu_0DE+t_{10}E-t_{00}D\cr
\mu_0DE+\nu_1E^2+\la_0ED+t_{01}E-t_{10}E\cr}\ ,
\eea
where
\bea
{\cal H}_1&=&\left(\matrix{
-\G^{110}_{101}-\G^{110}_{011} &\G^{101}_{110} &\G^{011}_{110} \cr
\G^{110}_{101} &-\G^{101}_{110}-\G^{101}_{011} &\G^{011}_{101} \cr 
}\right)\nn
{\cal H}_2&=&\left(\matrix{
-\G^{100}_{010}-\G^{100}_{001} &\G^{010}_{100} &\G^{001}_{100} \cr
\G^{100}_{010} &-\G^{010}_{100}-\G^{010}_{001} &\G^{001}_{010} \cr 
}\right)\ .
\eea

In order to demonstrate that there exist solutions to these equations
we will prove the existence of a one-dimensional representation for
the special choice of boundary conditions
\bea
(h_1)^{00}_{11}&=&-\a = - (h_1)^{00}_{00}\nn
(h_L)^{11}_{00}&=&-\b = - (h_L)^{11}_{11}\ ,
\eea
which correspond to injection of particles at sites $1$ and $2$ with
probability $\a$ if both sites are empty, and extraction of particles
at sites $L-1$ and $L$ with probability $\b$ if both sites are
occupied. 
The one-dimensional representation exists if $D=E=1$ and
\bea
\a&=&\b=\G^{110}_{101}+\G^{110}_{011}-\G^{101}_{110}-\G^{101}_{110}
=\G^{100}_{010}+\G^{100}_{001}-\G^{010}_{100}-\G^{001}_{100}\ ,\nn
\G^{110}_{101}+\G^{011}_{101}&=&\G^{101}_{011}+\G^{101}_{110}
\ ,\quad 
\G^{100}_{010}+\G^{001}_{010}=\G^{010}_{001}+\G^{010}_{100}\ .
\eea
The corresponding probability distribution is trivial
\be
P_L(\tau_1\ldots\tau_L)=\frac{1}{2^L}\ ,
\ee
which means that all configurations are equally represented.
Although the existence of the 1-d representations is a nontrivial fact,
the corresponding physics is not particularly interesting. It would be
very interesting to construct finite or infinite dimensional 
representation and use them to compute currents, density profiles 
{\sl etc}.

\vspace*{.5cm}
\appendix{\Large\bf Appendix C}
\vspace*{.5cm}

In this appendix we consider the special case $0=\gamma=\delta$ and
$\a=\b=p-q$ of the algebra \r{de}. As was shown in \r{qos} the
quadratic algebra reduces to a Q-osciallator algebra. We introduce the
following notation for Q-numbers 
\be
\{n\}=\frac{1-Q^n}{1-Q}\ ,
\ee
and Q-binomials
\be
{\cal C}^p_n(Q) = \frac{\{n\}\{n-1\}\dots
\{2\}}{\{n-p\}\{n-p-1\}\dots\{2\}\{p\}\{p-1\}\dots\{2\}}=
\frac{\prod_{r=1}^n(1-Q^r)}{\prod_{s=1}^p(1-Q^s)\prod_{t=1}^{n-p}(1-Q^t)}
\ . 
\ee
In order to compute the current or correlation functions a convenient
representation of $C^N=(D+E)^N$ is required. In terms of $a$ and
$a^\dagger$ we find
\be
C^N=\frac{1}{p^N}\left(\frac{2}{1-Q}+\frac{a+a^\dagger}{\sqrt{1-Q}}
\right)^N=\left(\frac{2}{p-q}\right)^N\sum_{j=0}^N {\cal C}^j_N(1)
(a+a^\dagger)^j \left(\frac{\sqrt{1-Q}}{2}\right)^j\ .
\ee
In order to evaluate {\sl e.g.} the normalization $\langle
0|C^L|0\rangle$ it is convenient to decompose powers of $a+a^\dagger$
into normal-ordered expressions defined {\sl via}
\be
:(a+a^\dagger)^n:=\sum_{p=0}^n{\cal C}^p_n(Q) {a^\dagger}^p a^{n-p}\ .
\ee
The decomposition is of the form
\be
(a+a^\dagger)^n=\sum_{m=0}^{\left[\frac{n}{2}\right]}
M^{(m)}_n :(a+a^\dagger)^{n-2m} :\ .
\ee
Using the identity
\be
(a+a^\dagger)\ :(a+a^\dagger)^n: = :(a+a^\dagger)^{n+1}: +\{n\}
:(a+a^\dagger)^{n-1}:\ ,
\ee
one readily obtains recursive expressions for $M^{(k)}_n$
\be
M^{(0)}_n=1\ ,\ 
M^{(1)}_n=\sum_{l=1}^{n-1}\{l\}\ ,\ 
M^{(2)}_n=\sum_{l=1}^{n-3}\{l\}M^{(1)}_{l+2}\ ,\ 
M^{(3)}_n=\sum_{l=1}^{n-5}\{l\}M^{(2)}_{l+4}\ ,\ldots 
\label{mkn}
\ee
We first note the result for the cases $Q=1$ (no deformation) and $Q=0$
\bea
M^{(m)}_n\bigg|_{Q=1}&=&\pmatrix{n\cr 2m}\prod_{k=0}^{m-1}(2k+1)
=\frac{n!}{(n-2m)!\ (2m)!!}\ ,\nn
M^{(m)}_n\bigg|_{Q=0}&=&\pmatrix{n-1\cr m}-\pmatrix{n-1\cr m-2}\ .
\label{q1}
\eea

After some tedious computations we find for arbitrary $Q$
\bea
M^{(1)}_n&=&\frac{n-\{n\}}{1-Q}\ ,\nn
M^{(2)}_n&=&\frac{1}{(1-Q)^2}\left[n\left(\frac{n-1}{2}-\{n-2\}\right)
-\{n\}\left(\frac{\{n-3\}}{\{2\}}-\{n-3\}\right)\right]\ .
\label{q2}
\eea
We did not succeed in obtaining a closed form for $M^{(k)}_n$ for
general $k$. The difficulty can be traced back to the recurrence
relation \r{mkn} which although it is written entirely in terms of
$Q$-numbers does not have $Q$-number solutions (the sum of $Q$-numbers
is not a $Q$-number).

\vspace*{.5cm}
\appendix{\Large\bf Appendix D}
\vspace*{.5cm}

In this appendix we would like to show that the DEHP-Ansatz gives a
new way to construct irreducible representations of the quantum group
$U_q(SU(2))$ and that the quantum plane appears in a natural way.
We consider the special case $0=\xi=\eta$, $\la=q/p$,
$z_2={p}/({p-q})$ of \r{de}. The boundary conditions are taken such
that $0=\gamma=\delta$, then the limit $\a\rightarrow 0$,
$\b\rightarrow 0$ is performed. For $0=\a=\b=\g=\d$ the XXZ quantum
spin hamiltonian \r{A5d} corresponding to the diffusion process is
invariant under the quantum algebra $U_q(SU(2))$ (\cite{ps,vlad1} and
references therein). The stationary state \r{B1} 
is no more unique since the ground state of the ferromagnetic chain is
$L+1$-times degenerate corresponding to a multiplet of $sl_q(2)$.
As the boundary conditions \r{de} become ill-defined in
the case $0=\a=\b=\g=\d$ we carefully take the limit $\a\rightarrow 0$,
$\b\rightarrow 0$ with $0=\g=\d$ as follows
\bea
\a&=&\frac{\e}{e},\ \b=\frac{\e}{d},\ {\bar D}=\e D,\ {\bar E}=\e
E,\nn
\w {\bar E} &=& e\w,\ {\bar D}\v = d\v,\ 
\eea
where ${\bar D}$ and ${\bar E}$ are seen to obey the quantum plane
equation 
\be
{\bar D}{\bar E}={\cal Q}^2{\bar E}{\bar D}\ ,\ {\cal Q}^2=\frac{q}{p}\ .
\ee
The quantum-mechanical state corresponding to the stationary state of
the diffusion process is obtained by applying \r{qmstate}
\be
|0\rangle=\frac{1}{{\cal N}}\sum_{k=0}^Ld^{L-k}e^k\sum_{i_1<i_2<\ldots<
i_k}{\cal Q}^{2(i_1+i_2+\ldots +i_k -
\frac{k+k^2}{2})}\prod_{l=1}^k\sigma^-_{i_l} |\up\up\ldots\up\rangle ,
\ee
where $\frac{1}{{\cal N}}$ is a normalization factor. Note that
$|0\rangle$ is a linear combination of $L+1$ independent wave functions.
In order to get the ground state of the XXZ chain \r{A5d} we still
have to perform the similarity transformation \r{sim}, which changes
the probability distribution of the stationary state \r{B1} into
\be
\w\prod_{i=1}^L\left(\tau_i\La{\cal Q}^{i-1} D+(1-\tau_i) E\right)\v = 
\La^L{\cal Q}^{\frac{L(L-1)}{2}} \w\prod_{i=1}^L\left(\tau_i
D+\frac{(1-\tau_i)}{\La{\cal Q}^{i-1}} E\right)\v\ .
\ee
This yields the following result for the similarity-transformed state
$|0\rangle_U$, which is the ground state of the $U_q(SU(2))$-invariant
XXZ hamiltonian \r{A5d}
\be
|0\rangle_U=\frac{1}{{\cal N}}\La^L{\cal
Q}^{\frac{L(L-1)}{2}}\sum_{k=0}^L
\La^{-k}d^{L-k}e^k\sum_{i_1<i_2<\ldots< i_k}{\cal Q}^{i_1+i_2+\ldots
+i_k - k^2}\prod_{l=1}^k\sigma^-_{i_l} |\up\up\ldots\up\rangle . 
\ee
We note that the k'th term in the sum is proportional to the state
$(S^-)^k|\up\up\ldots\rangle$ obtained by acting with the
quantum-group generators  
\be
S^-=\sum_{l=1}^L{\cal Q}^{\frac{1}{2}\sum_{k=1}^{l-1}\sigma^z_k}\
\sigma^-_l\ {\cal Q}^{-\frac{1}{2}\sum_{m=l+1}^{L}\sigma^z_m}\ .
\ee
We believe that using the DEHP Ansatz in order to get irreducible
representations of quantum groups may have other applications.
\vspace*{.5cm}
\appendix{\Large{\bf Appendix E}}
\vspace*{.5cm}

Here we address the problem of existence of solutions to the system of
relations \r{h1l},\r{bcs},\r{diffalg1}-\r{diffalg3}. 
It is convenient to define symmetric and antisymmetric combinations of
rates 
\bea
\G_{a}^\pm&=&\G^{01}_{11}\pm\G^{10}_{11}\ ,\
\G_{b}^\pm=\G^{01}_{00}\pm\G^{10}_{00}\ ,\ 
\G_{c}^\pm=\G^{01}_{10}\pm\G^{10}_{01}\ ,\nn
\G_{d}^\pm&=&\G^{11}_{10}\pm\G^{11}_{01}\ ,\
\G_{e}^\pm=\G^{00}_{10}\pm\G^{00}_{01}\ ,\ \Delta=
\frac{\a+\d}{\b+\g}\ .
\label{newvar}
\eea
Note that due to positivity of the rates all symmetric combinations
are automatically positive. 
With
\be
E=\frac{\b+\g}{\a\b-\g\d}\ ,\quad D=\frac{\a+\d}{\a\b-\g\d}
\ee
it is straightforward to show that 1-d representations exist under the
condition that 

\bea
\G_{a}^++\G_{b}^+&=&\G_{d}^+\Delta +\G_{e}^+\Delta^{-1},\nn 
\Delta(\G^{11}_{00}+\G_{d}^+)&=&\G^{00}_{11}\Delta^{-1}+\G_{a}^+\ ,\nn
\G_{e}^-\Delta^{-1}+\Delta\G_{d}^-+2\G_{c}^-+\G_{a}^-+\G_{b}^-&=&
-\frac{2}{E}(1+\Delta^{-1})\ .
\label{exi}
\eea
It is obvious that one can choose the $12$ rates such that equations
\r{exi} are satisfied.


\vspace*{.5cm}
\appendix{\Large{\bf Appendix F:} Mean-Field Analysis}
\vspace*{.5cm}

In this appendix we give a summary of the mean-field analysis for the
partially asymmetric diffusion process on a lattice with $L$ sites.
Our discussion follows \cite{ddm,md}, which deals with the completely
asymmetric case.
The particles hop with rate $p$ ($q$) to the right (left) and are
injected (extracted) with rates $\a$ ($\g$) at site 1 and rate $\d$
($\b$) at site $L$.

In a stationary state the density at site $j$ is time-independent,
which implies
\be
0=\frac{d\t{j}}{dt}= (q-p)\2t{j-1}{j}+(p-q)\2t{j}{j+1}+p\t{j-1}
+q\t{j+1}-(p+q)\t{j}\ .
\ee
Denoting $\langle\tau_j\rangle$ by $t_j$ and decoupling the two-point
functions $\2t{j}{k} = t_jt_k$ leads to the following set of
mean-field equations 
\bea
pt_{j-1}+qt_{j+1}-pt_{j-1}t_j-qt_jt_{j+1}&=&
(p+q)t_{j}-pt_{j}t_{j+1}-qt_jt_{j-1}\ ,\quad j=2\ldots L-1\label{mfbulk}\\
\a(1-t_1)+qt_2(1-t_1)&=&\g t_1+pt_1(1-t_2)\ ,\nn
\b t_L+qt_L(1-t_{L-1})&=&\d (1-t_L)+pt_{L-1}(1-t_L)\ .
\label{mf}
\eea
The bulk equations \r{mfbulk} can be rewritten as
\be
t_{j+1}t_j = -\frac{q}{p-q}t_{j+1}+\frac{p}{p-q}t_j+c\ ,
\label{mfrec}
\ee
where $c$ is an arbitrary constant (related to the current $J$).
The net mean-field current from site $j$ to site $j+1$ is defined as
\be 
J=pt_j(1-t_{j+1})-q(1-t_j)t_{j+1}=-(p-q)c\ ,
\label{mfJ}
\ee
and is independent of position as it should be for a stationary state.
If we define
\be
s_j=\frac{p-q}{p+q}\left(t_j+\frac{q}{p-q}\right)\ ,
\ee
the new quantities $s_j$ are seen to obey the recursion
\bea
s_{j+1}&=&1-\frac{c'}{s_j}\ ,\quad
c'=\frac{pq}{(p+q)^2}-\left(\frac{p-q}{p+q}\right)^2 c\ ,\nn
s_1&=&\frac{p\a+q\g+pq-(p+q)^2c'}{(p+q)(\a+\g)}\ ,\ 
s_L=\frac{p\d+q\b-pq+(p+q)^2c'}{(p+q)(\b+\d)}\ .
\label{mfs}
\eea
We note that $0\leq c'$ as otherwise $s_2>1$, which is
unphysical as $0\leq t_j\leq 1\ ,\ \forall j$ (which implies
$\frac{q}{p+q}\leq s_j\leq\frac{p}{p+q}$). Like for the completely
asymmetric case we now have to distinguish three cases:
\begin{itemize}
\item{} $c' > \frac{1}{4}$

In this case the recursion has no fixed point and $s_j$ would
eventually turn negative, which leads to an unphysical solution. Thus
we can exclude this case.

\item{} $c' = \frac{1}{4}$

This case corresponds to the maximal current phase $C$ in the phase
diagram (see Fig.2). The current is given by (see above)
\be
J=-(p-q)c=\frac{(p+q)^2}{(p-q)}c'-\frac{pq}{p-q}=\frac{p-q}{4}\ ,
\label{jp}
\ee
which is the same as the exact result \r{C11}.

\item{} $c' < \frac{1}{4}$

In this case the recursion \r{mfs} has two fixed points
\be
s_{\pm}=\frac{1}{2}\left(1\pm\sqrt{1-4c'}\right)\ .
\label{fixp}
\ee
Writing $s_j=\sqrt{c'}\frac{u_{j+1}}{u_j}$ we see that the $u_j$'s are
subject to the recursion
\be
u_{j+1}+u_{j-1}=\frac{1}{\sqrt{c'}}u_{j}\ ,
\label{che}
\ee
which is recognized as a special case of the recursion relation for
Chebyshev polynomials. Eqn \r{che} is solved formally as
\be
u_n(\theta)=\sin[(n-1)\theta +\phi]\ ,
\theta=\arccos\left(\frac{1}{2\sqrt{c'}}\right)\ ,
\ee
(note that $\theta$ is complex) which leads to the following
expression for $s_n,\ j=n\ldots L$ 
\be
s_n=\sqrt{c'}\frac{\sin[n\theta +\phi]}{\sin[(n-1)\theta +\phi]}\ .
\label{che2}
\ee
Using the two boundary conditions \r{mfs} in \r{che2} completely fixes
the values of $\phi$ and $c'$ as functions of $\a$, $\b$, $\g$, $\d$,
$p$ and $q$. For simplicity we introduce the notation $s_1=d_1-d_2c'$
and $s_L=d_3+d_4c'$. The boundary condition for $s_1$ implies that
\be
\cot(\phi)=\frac{2(2d_1-1)\cos^2(\theta)-d_2}{\sin(2\theta)}\ ,
\label{phi1}
\ee
whereas the one for $s_L$ yields
\be
\cot(\phi)=-\frac{2\cos(\theta)\cos(L\theta)-
(d_4+4d_3\cos^2(\theta))\cos[(L-1)\theta]}{2\cos(\theta)\sin(L\theta)-
(d_4+4d_3\cos^2(\theta))\sin[(L-1)\theta]}\ .
\label{phi2}
\ee
Equating \r{phi1} and \r{phi2} we obtain
\bea
0&=&d_1(1-d_3)\sin[(L+3)\theta]+
[(2d_1-d_2-1)(1-d_3)+2d_1(1-2d_3-d_4)]\sin[(L+1)\theta] \nn
&&+[(2d_1-d_2-1)(1-2d_3-d_4)+(d_1-1)(1-d_3)-d_1d_3]\sin[(L-1)\theta] \nn
&&-[(2d_1-d_2-1)d_3+(1-d_1)(1-2d_3-d_4)]\sin[(L-3)\theta]\nn
&&+ d_3(1-d_1)\sin[(L-5)\theta]\ .
\label{sins}
\eea
In the large-$L$ limit \r{sins} turns into a fourth order polynomial
equation in $z=\exp(2i\theta)$ ($\theta$ is complex). The polynomial
equation can then be solved explicitly for $\theta$ (and thus $c'$) as
a function of $\a,\b,\g,\d,p,q$. \vskip .5cm

However, there exists a much simpler way to
determine the mean field current \cite{ddm}: in the low-density phase B
we start out infinitesimally close to the unstable fixpoint $s_-$,
{\sl i.e.} $s_1=s_-+\varepsilon$. The density stays at $s_-$
throughout the bulk and only deviates towards $s_+$ at the right end
of the chain. Using the fact that $c'=s_-(1-s_-)$ in the expression
\r{mfs} for $s_1$, and then setting $s_1=s_-$ immediately
yields $s_-$, and thus also $c'$, as a function of $p$, $q$, $\a$ and
$\g$. Inserting the resulting expression for $c'$ into \r{jp} then 
yields the current as a function of $\a$, $\g$, $p$ and $q$. The
result found to be identical to the exact expression \r{C10}. 

An analogous analysis can be carried out in the high-density phase $A$.
Again the result is identical to the exact expression \r{C9}
\end{itemize}

We also can use mean-field theory to determine the density profile in
phases $A$ and $B$. From our discussion above it is clear that in
phase $B$ the density profile in the bulk is essentially constant and
equal to the value of the unstable fixed point (we switch back from
the $s_j$-variables to $t_j$ variables)
\be
t_-=\frac{1}{2}(1-\sqrt{1+4c})=\frac{1}{2}(1-\sqrt{1-\frac{4J_B}{p-q}})\
. 
\ee
Analogously, in phase $A$ the profile in the bulk is constant and
equal to the value at the stable fixed point $t_+$
\be
t_+=\frac{1}{2}(1+\sqrt{1-\frac{4J_A}{p-q}})\
. 
\ee
Using the expression for the currents these values can be determined
explicitly 
\bea
t_-=:m_{<,MF}&=&\frac{1}{2}-\frac{1}{p-q}
\sqrt{\frac{(p-q)^2}{4}-\a(\a+\g)\kappa_+(\a,\g)+\g[\a+\g+p-q]}\ ,\nn
t_+=:m_{>,MF}&=&\frac{1}{2}+\frac{1}{p-q}
\sqrt{\frac{(p-q)^2}{4}-\b(\b+\d)\kappa_+(\b,\d)+\d[\b+\d+p-q]}\ .
\eea
These expressions coincide with \r{magn}. Finally we note that the
mean-field result for the correlation length $\zeta$ does not
repreduce \r{corrl} and is incorrect.

\end{document}